\documentclass[twocolumn]{aastex62}

\hypersetup{linkcolor=red}
\newcommand\st{\bgroup\markoverwith
{\textcolor{magenta}{\rule[0.5ex]{2pt}{1pt}}}\ULon}

\usepackage{amsmath, bm, hyperref, ulem}
\usepackage{xcolor,colortbl, array, makecell} 
\newcolumntype{?}[1]{!{\vrule width #1}}
\definecolor{Gray}{gray}{0.85}
\definecolor{LightCyan}{rgb}{0.88,1,1}
\usepackage{savesym} \savesymbol{tablenum}
\usepackage[detect-weight=true, detect-family=true]{siunitx} \restoresymbol{SIX}{tablenum}
\usepackage[T1]{fontenc}
\usepackage[overload]{textcase}
\graphicspath{{./}{Figures/}}

\accepted{14 September 2020 by AJ}

\shorttitle{ACCESS: WASP-31\MakeLowercase{b}} \shortauthors{McGruder et al.}

\begin{document} 
\title{ACCESS: Confirmation of no potassium in the atmosphere of WASP-31b} 

\correspondingauthor{Chima D. McGruder} \email{chima.mcgruder@cfa.harvard.edu}

\author[0000-0002-6167-3159]{Chima D. McGruder}\altaffiliation{NSF Graduate Research Fellow}\affiliation{Center for Astrophysics ${\rm \mid}$ Harvard {\rm \&} Smithsonian, 60 Garden St, Cambridge, MA 02138, USA}

\author[0000-0003-3204-8183]{Mercedes L\'opez-Morales} \affiliation{Center for Astrophysics ${\rm \mid}$ Harvard {\rm \&} Smithsonian, 60 Garden St, Cambridge, MA 02138, USA}

\author[0000-0001-9513-1449]{N\'estor Espinoza} \affiliation{Space Telescope Science Institute (STScI), 3700 San Martin Dr, Baltimore, MD 21218}

\author[0000-0002-3627-1676]{Benjamin V. Rackham}
\altaffiliation{51 Pegasi b Fellow}
\affiliation{Department of Earth, Atmospheric and Planetary Sciences, and Kavli Institute for Astrophysics and Space Research, Massachusetts Institute of Technology, Cambridge, MA 02139, USA}

\author[0000-0003-3714-5855]{D\'aniel Apai}\affiliation{Department of Astronomy/Steward Observatory, The University of Arizona, 933 N. Cherry Avenue, Tucson, AZ 85721, USA}\affiliation{Lunar and Planetary Laboratory, The University of Arizona, 1640 E. Univ. Blvd, Tucson, AZ 85721}\affiliation{Earths in Other Solar Systems Team, NASA Nexus for Exoplanet System Science}

\author[0000-0002-5389-3944]{Andr\'es Jord\'an}\affiliation{Facultad de Ingenier\'ia y Ciencias, Universidad Adolfo Ib\'a\~nez, Av.\ Diagonal las Torres 2640, Pe\~nalol\'en, Santiago, Chile}\affiliation{Millennium Institute for Astrophysics, Santiago, Chile}

\author{David J. Osip}\affiliation{Las Campanas Observatory, Carnegie Institution of Washington, Colina el Pino, Casilla 601 La Serena, Chile}

\author[0000-0003-4157-832X]{Munazza K. Alam}\altaffiliation{NSF Graduate Research Fellow}\affiliation{Center for Astrophysics ${\rm \mid}$ Harvard {\rm \&} Smithsonian, 60 Garden St, Cambridge, MA 02138, USA}

\author[0000-0003-2831-1890]{Alex Bixel}\affiliation{Department of Astronomy/Steward Observatory, The University of Arizona, 933 N. Cherry Avenue, Tucson, AZ 85721, USA}\affiliation{Earths in Other Solar Systems Team, NASA Nexus for Exoplanet System Science}

\author[0000-0002-9843-4354]{Jonathan J. Fortney}\affiliation{Department of Astronomy and Astrophysics, University of California, Santa Cruz, USA}

\author[0000-0003-4155-8513]{Gregory W. Henry}\affiliation{Center of Excellence in Information Systems, Tennessee State University, Nashville, TN 37209, USA}

\author[0000-0002-4207-6615]{James Kirk}\affiliation{Center for Astrophysics ${\rm \mid}$ Harvard {\rm \&} Smithsonian, 60 Garden St, Cambridge, MA 02138, USA}

\author[0000-0002-8507-1304]{Nikole K. Lewis}\affiliation{Department of Astronomy and Carl Sagan Institute, Cornell University, 122 Sciences Drive, 14853, Ithaca, NY, USA}

\author[0000-0003-0650-5723]{Florian Rodler}\affiliation{European Southern Observatory, Alonso de Cordova 3107, Vitacura, Santiago de Chile, Chile}  

\author[0000-0001-6205-6315]{Ian C. Weaver}\affiliation{Center for Astrophysics ${\rm \mid}$ Harvard {\rm \&} Smithsonian, 60 Garden St, Cambridge, MA 02138, USA}


\begin{abstract} 
We present a new optical (400--950\,nm) transmission spectrum of the hot Jupiter WASP-31b (M=0.48\,M\textsubscript{J}; R= 1.54\,R\textsubscript{J}; P=3.41\,days), obtained by combining four transits observations. These transits were observed with IMACS on the Magellan Baade Telescope at Las Campanas Observatory as part of the ACCESS project. We investigate the presence of clouds/hazes in the upper atmosphere of this planet as well as the contribution of stellar activity on the observed features. In addition, we search for absorption features of the alkali elements Na I and K\,I, with particular focus on K\,I, for which there have been two previously published disagreeing results. Observations with HST/STIS detected K\,I, whereas ground-based low- and high-resolution observations did not. We use equilibrium and non-equilibrium chemistry retrievals to explore the planetary and stellar parameter space of the system with our optical data combined with existing near-IR observations. Our best-fit model is that with a scattering slope consistent with a Rayleigh slope ($\alpha = 5.3^{+2.9}_{-3.1}$), high-altitude clouds at a log cloud top pressure of -3.6$^{+2.7}_{-2.1}$ bars, and possible muted  H\textsubscript{2}O features. We find that our observations support other ground-based claims of no K\,I. Clouds are likely why signals like H\textsubscript{2}O are extremely muted and Na or K cannot be detected. We then juxtapose our Magellan/IMACS transmission spectrum with existing VLT/FORS2, HST/WFC3, HST/STIS, and \textit{Spitzer} observations to further constrain the optical-to-infrared atmospheric features of the planet. We find that a steeper scattering slope ($\alpha = 8.3 \pm 1.5$) is anchored by STIS wavelengths blueward of 400\,nm and only the original STIS observations show significant potassium signal.
\end{abstract}



\section{Introduction} \label{sec:intro} 
We are at the forefront of an exoplanet revolution with \added{ over 4200 planets discovered to date. Of these planets, over half have their fundamental properties} (i.e.\ orbital parameters, masses and radii) characterized, which give a basic understanding of their formation and interior composition \footnote{see, e.g., \url{https://exoplanets.nasa.gov/exoplanet-catalog/}}. However, to obtain a more in-depth understanding of how planets form and evolve, it is necessary to study their atmospheres in detail. 


To date, transmission spectroscopy is the most successful approach to probe exoplanets' atmospheres. However, most of the instruments currently used for transmission spectroscopy were not designed with this science in mind, and unconstrained instrumental systematics are a major limitation to data precision. In the case of ground-based observations, effects related to the Earth's atmosphere are also a non-trivial factor. Because each telescope/instrument combination has its own unique systematics, there is no specific prescription to handle all of these limitations, which can lead to different interpretations of the same data based solely on the detrending process. This problem has caused several claims and counterclaims in this field \added{(e.g. \citeauthor{Espinoza2019} \citeyear{Espinoza2019} vs. \citeauthor{Sedaghati2017} \citeyear{Sedaghati2017}; \citeauthor{Diamond-Lowe2018} \citeyear{Diamond-Lowe2018} vs. \citeauthor{Southworth2017} \citeyear{Southworth2017})}. WASP-31b is one such planet that has had a detection of K\,I \citep{Sing2015WASP31b} that was later refuted \citep{Gibson2017WASP31, Gibson2019WASP31}. To exacerbate the situation, stellar activity can mimic planetary features \added{\citep[see e.g.,][]{Pont:2013, McCullough2014, Espinoza2019}}, and tellurics might mask signals in ground-based observations \citep{Gibson2017WASP31}. Thus, the best approach in understanding the atmospheric transmission spectrum of an exoplanet is to repeatedly observe the planet and robustly detrend the data.

\added{\object{WASP-31} is a low-activity (3-5 Gyr), F6V dwarf, with low metallicity ([Fe/H] = -0.19)} and an effective temperature of about 6300\,K. WASP-31 (V = 11.7) is also a visual binary with a V $\sim$ 15.8 mag companion approximately $\SI{35}{\arcsecond}$ away. \object{WASP-31b} is a transiting low-density ($\rho$ = 0.129 $\rho$\textsubscript{J}), inflated (R = 1.54 R\textsubscript{J}) hot Jupiter with an orbital radius of 0.047 AU, a period of 3.4 days, a mass of 0.48 M\textsubscript{J}, and a zero bond albedo equilibrium temperature of 1575 K \citep{Anderson2011_W31discov}. This planet is a particularly good target for transmission spectroscopy because its lower gravity leads to a larger atmospheric scale height and stronger signals \deleted{\citep{Kreidberg2017}}.

WASP-31b's transmission spectrum was observed with the Space Telescope Imaging Spectrograph (STIS) and Wide Field Camera 3 (WFC3) instruments on the Hubble Space Telescope (HST). These observations were combined with \textit{Spitzer} photometry to construct a high signal-to-noise transmission spectrum. Analysis of those data indicated a Rayleigh scattering signature at short wavelengths, a cloud-deck at longer wavelengths, a strong potassium absorption feature (4.2$\sigma$ confidence, with 78\,{\AA} wide bins), but no sodium \citep{Sing2015WASP31b}. This is mostly consistent with ground-based observations later made with FORS2 on the VLT, aside from the detection of K\,I \citep{Gibson2017WASP31}. \citeauthor{Gibson2017WASP31} re-analyzed the HST observations using Gaussian processes (GPs) to correct for systematics and found that the potassium detection significance decreased to 2.5$\sigma$. Additional high-resolution observations (R > 80000) with the VLT's UV-Visual Echelle Spectrograph (UVES) did not detect sodium or potassium, supporting the idea that HST observations also might be susceptible to spurious signals produced by systematics \citep{Gibson2019WASP31}. However, because the detection of the K\,I feature from the ground can be contaminated by Earth’s O\textsubscript{2} telluric features, it is worth further investigating these results.

Here we present four new transit observations of WASP-31b with spectral resolutions 300 < R < 1300 and wavelength coverage between 0.4--0.97$\mu$m, obtained between 2013 and 2019 as part of ACCESS\footnote{ACCESS is a multi-institutional collaboration aiming to produce a large, homogeneous library of optical spectra of exoplanet atmospheres. \url{http://project-access.space/}}. ACCESS utilizes the multi-object spectrograph (MOS) IMACS \citep{2011Dressler} on the 6.5-meter Magellan Baade Telescope at Las Campanas Observatory, which allows for simultaneous collection of spectra from the exoplanet host star and many comparison stars to enable better systematic corrections. To date, ACCESS has produced transmission spectra of WASP-6b \citep{Jordan:2013}, GJ~1214b \citep{rackham2017}, WASP-4b \citep{Bixel2019}, WASP-19b \citep{Espinoza2019}, and WASP-43b \citep{weaver2019access}. To these we add here optical measurements of WASP-31b, which we combine with previously published HST/WFC3 and \textit{Spitzer} IR data to produce a new transmission spectrum with coverage between 0.4 and 5.0 $\mu$m.

This paper is structured as follows: In Section~\ref{sec:obs}, we present our Magellan/IMACS observations of WASP-31b in addition to photometric monitoring observations to constrain stellar variability. In Section~\ref{sec:data_red_lc_analysis}, we discuss our transit light curve extraction process, binning schemes, and detrending techniques. We outline the production of our final transmission spectrum and present a qualitative comparison of this spectrum to data from the literature in Section~\ref{sec:trans_spec}. In Section~\ref{sec:atmo_retriev}, we introduce the retrievals used to analyze the transmission spectra and report our best-fit results compared to best fits obtained from previous data. Interpretations of the retrieval results in context of WASP-31b's overall atmospheric properties are discussed in Section~\ref{sec:retriev_interp}. We conclude and summarize in Section~\ref{sec:conclusion}.

\section{Observations} \label{sec:obs} 
\subsection{General Setup} \label{sec:gen_setup}
We observed four transits of WASP-31b on UT dates UT130226, UT130425, UT140222, and UT190314 (UTYYMMDD) with IMACS on the 6.5-meter Baade Magellan Telescope at Las Campanas Observatory in Chile. \added{The large time-lapse between the first three transits and the last, is because transit UT130226 could not be used for the transmission spectrum (as discussed in appendix \ref{appx:atmo_retriev_130226}), which we only found to be the case later in the analysis process.} 

\begin{figure}[htb]
    \centering
    \includegraphics[width=\linewidth]{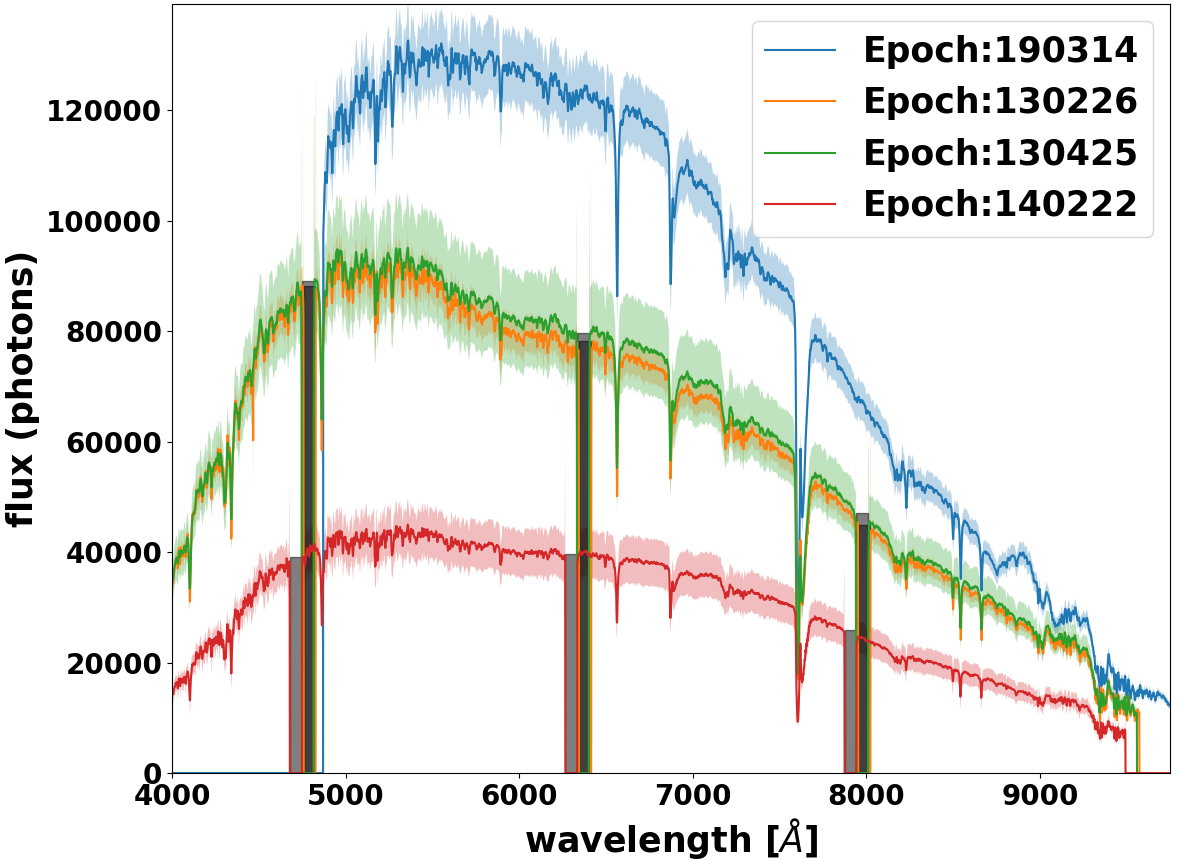}
    \caption{Median extracted spectra of WASP-31 from transits UT130226 (orange), UT130425 (green), UT140222 (red), and UT190314 (blue). The shaded regions of the same color extending past the median lines are the 1-$\sigma$ range of counts extracted for that night. The difference in counts between the f/4 observations are because of the exposure times. Different throughputs with the f/4 and f/2 setups explain the different shape of the spectra from UT190314 and from the f/4 observations. Dark shaded regions indicate chip gaps; as can be seen there are 3 gaps for f/4 observations and usable data starts at .48 $\mu$m for the f/2 mode.}
    \label{fig:finExtSpec} 
\end{figure}

IMACS is a wide-field imager and spectrograph with two 8Kx8K CCD mosaic cameras at the f/2 and f/4 foci \citep{2011Dressler}. Each camera has a multi-object spectroscopy (MOS) mode, allowing for custom slit masks designs. The f/2 camera has a large, \SI{27.4}{\arcmin} field of view (FoV) and resolving powers up to R $\sim$ 5000. The f/4 camera has a smaller FoV (\SI{15.4}{\arcmin}x\SI{15.4}{\arcmin}) but higher spectral resolution (max R $\sim$ 9000). The orientation of the f/4 camera causes each spectrum to be dispersed over up to 4 of the 8 CCDs, whereas in the f/2 mode each spectrum is dispersed over up to two of the CCDs. The gaps between CCDs introduce gaps in the spectra, as shown in Figure \ref{fig:finExtSpec}. Our first three transits (UT130226, UT130425, and UT140222) were observed with the f/4 camera and the last one (UT190314) with the f/2 camera.

The large FoV of IMACS allows for simultaneous observations of the target and several comparison stars to correct for common time-series variations. We imposed strict criteria for the comparison star selection to ensure they are photometrically stable, have similar apparent magnitudes, and similar spectral types (color distance < 1\footnote{color distance, D, is defined in \cite{rackham2017} as D = $\sqrt{[(B-V)_c - (B-V)_t]^2 + [(J-K)_c - (J-K)_t]^2}$ where $t$ and $c$ represent the target and comparison star magnitudes, respectively.}), as described in detail in \cite{rackham2017}. The comparison stars selected for each observation are summarized in Table \ref{tab:comp_stars}.

We designed masks with slits wide enough to prevent slit losses in typical seeing conditions at LCO but narrow enough to prevent contamination from nearby stars. This resulted in slit widths of $\SI{10}{\arcsecond}$ and $\SI{5}{\arcsecond}$ for the f/2 and f/4 mode, respectively. The lengths of the slits were $\SI{22}{\arcsecond}$ for f/2 and $\SI{12}{\arcsecond}$ for f/4 to adequately sample the sky background. We used a 300~line/mm grating at blaze angle of 4.3$^{\circ}$ for the f/4 observations and a 300~line/mm grism at blaze angle of 17.5$^{\circ}$ for the f/2 observation. These setups provided a wavelength coverage of 0.4--0.95\,$\mu$m and 0.45--0.97\,$\mu$m for the f/4 and f/2 observations, respectively. Finally, to reduce readout time and improve the duty cycle of the observations, we binned the detectors by 2$\times$4 (dispersion $\times$ spatial directions) for the UT130266 transit. \added{Transit UT130266 was the first observation done as part of ACCESS, while we were still experimenting with settings. After this first try we realized that seeing conditions at Las Campanas Observatory are often very good (typically 0.7 arcsec or better), so 2$\times$4 binning undersamples the spectra’s PSF. We decided to use 2$\times$2 binning for all subsequent WASP-31b observations, as an extra precaution to ensure proper sampling of the spectrum’s PSF under all seeing conditions.} Given the wide slits, the spectral resolution of the observations is seeing-limited, with an average resolution of R $\sim$ 900 (FWHM dispersion of 8.08{\AA}). 

\begin{deluxetable}{ccccccc}[htb]
    \caption{Target and comparison star magnitudes and coordinates from the UCAC4 catalog \citep{2013UCAC4}. We used comparisons 2--4 for the white-light fit of epoch UT130226, 1--4 for epochs UT130425 and UT140222, and stars 2, 5--7 for epoch UT190314. Comparison 1 for epoch UT130226 was saturated in most bins and not used for this transit.} 
    \label{tab:comp_stars}
    \tablehead{\colhead{Star} & \colhead{RA} & \colhead{Dec} & \colhead{B}
                              & \colhead{V} & \colhead{J} & \colhead{K}}
    \startdata 
  WASP-31 & 11:17:45.4 & -19:03:17.2 & 12.4 & 11.9 & 10.9 & 10.7 \\
        1 & 11:17:45.3 & -19:07:25.9 & 11.8 & 11.1 & 10.0 & 9.6 \\
        2 & 11:17:32.0 & -19:08:37.0 & 13.1 & 12.6 & 11.6 & 11.4 \\
        3 & 11:17:38.1 & -18:59:19.6 & 13.7 & 13.0 & 11.9 & 11.5 \\
        4 & 11:17:44.3 & -18:58:06.5 & 14.7 & 13.8 & 11.9 & 11.4 \\
        5 & 11:17:10.8 & -19:14:26.0 & 12.5 & 11.9 & 11.0 & 10.7 \\
        6 & 11:16:35.7 & -19:01:35.6 & 13.2 & 11.9 & 9.7 & 8.9 \\
        7 & 11:18:23.2 & -18:58:45.8 & 13.3 & 12.8 & 11.9 & 11.6
    \enddata 
\end{deluxetable}

\subsection{IMACS Data Collection} \label{sec:data_collect}
The transits of WASP-31b last 2.6 hours, so to cover the full transits and obtain enough baseline coverage to extract precise transit depths, we monitored the system during each transit epoch for about 5-hour continuous time blocks. We limited observations to epochs when WASP-31 was at airmasses below 1.7 during the full observation window to minimize atmospheric refraction effects in the data. Exposure times were adjusted from 27 to 60 seconds to provide maximum count rates of 25,000--35,000 in analog-to-digital units (ADU; gain= 0.56 e$^{-1}$/ADU for f/4 setup, 1.0 e$^{-1}$/ADU  for f/2 setup), which was well within the CCD's linearity limit \citep{Bixel2019}. More details of specific transit observations are given in Table~\ref{tab:ObsLog}.

For each night we collected bias frames, high-SNR quartz lamp flats, and calibration arcs with HeNeAr lamps, each with the same binning scheme as the science observations. From the biases we found that the levels were essentially constant across the detector, so we adopted a constant bias level on each science frame using the median of their corresponding overscan region. We did not take dark frames because dark counts are negligible for our exposure times \citep{rackham2017}. We took series of $\sim$20, 1 sec flat frames before and after the science frames with the same configuration as the science images. \added{The exposure times of the flats} were set to give a count rate of about 35,000 ADU or under in order to stay within the linear regime of the CCDs \citep{Bixel2019}. However, during the data reduction (see Section \ref{sec:data_red}), we found that flat-field corrections introduced additional noise in the data, similar to what was found with previous ACCESS observations \citep{rackham2017, Bixel2019, Espinoza2019, weaver2019access}. Therefore, we decided not to apply the flat-field correction to any of our data sets. Finally, we took exposures with He, Ne, and Ar lamps using calibration masks identical to the science masks but with 0.5$''$ slit widths to increase the spectral resolution and prevent saturation of the CCDs. These lamps were used to wavelength-calibrate the spectra, as described in Section \ref{sec:data_red}.

\begin{deluxetable*}{ccccccccc}[htb]
    \caption{Observing log for WASP-31b data sets from \textit{Magellan/IMACS}. The resolutions were calculated from the FWHM, which was estimated using the second moment method \citep{Markevich1989}. The average seeing conditions at LCO are $\SI{0.6}{\arcsecond} - \SI{0.7}{\arcsecond}$\citep{ThomasOsip2008} and is the resolution limiting factor.}
    \label{tab:ObsLog}
    \tablehead{{Transit Date} & {Obs. Start/} & {Airmass} & {Exposure} & {Readout +}& {Frames} &{binning} &{camera} &{min/max}\\ {(UTC)} & {End (UTC)} & & {Times (s)} & {Overhead (s)} & & & &{resolution}}
    \startdata 
    2013 Feb 27 & 02:58/8:20 & 1.02--1.3 & 40 & 29 & 283 & 2$\times$4 & f/4 & 719/1126 \\
    2013 Apr 25 & 23:06/5:05 & 1.02--1.46 & 40 & 26 & 321 & 2$\times$2 & f/4 & 472/822 \\ 
    2014 Feb 23 & 02:32/08:44 & 1.02--1.43 & 27--36 & 26 & 363 & 2$\times$2 & f/4 & 295/1169 \\
    2019 Mar 15 & 03:04/07:20 & 1.02--1.3 & 60 & 31 & 168 & 2$\times$2 & f/2 & 739/1233 \\
    \enddata 
\end{deluxetable*}

\subsection{Photometric Monitoring} \label{sec:PhotMon}
We analyzed photometric time series of the host star WASP-31 to constrain potential stellar contributions to the transmission spectrum of WASP-31b using observations from the Tennessee State University (TSU) Celestron 14-inch (C14) Automated Imaging Telescope (AIT) at Fairborn Observatory \citep{1999Henry_AIT}, the Transiting Exoplanet Survey Satellite \citep[TESS,][]{TESS2014}, and the All-Sky Automated Survey for Supernovae \citep[ASAS-SN,][]{2017Kochanek_ASAS-SN}. The combined time baseline of these observations cover all our transit epochs, as shown in Figure~\ref{fig:PhotMon}. However, the photometric precision varies between surveys, so we decided to analyze each observing season for each survey separately. 

\subsubsection{AIT Analysis}
TSU's AIT acquired 660 photometric observations through a Cousins R filter spanning 6 observing seasons (2011-12 through 2016-17). Standard deviations of a single observation from their seasonal means range between 3.2 and 4.2~mmag. This is near the limit of measurement precision with the AIT, as determined from the comparison stars in the field. The seasonal-mean differential magnitudes agree to a standard deviation of only 0.88 mmag, consistent with the absence of year-to-year variability and the low activity level of WASP-31. Periodogram analysis of each individual season resulted in the detection of significant periodicity only in the 2012--13 season.  We detected a small peak-to-trough amplitude of 4~mmag with a period of 8.60 
days, corresponding to the star's rotational period of 8.62 days as measured by \citep{Sing2015WASP31b}. This suggests WASP-31 is a relatively inactive star\footnote{See \cite{Sing2015WASP31b} for further details of AIT observations, reduction, and analysis.}. 

\subsubsection{TESS Analysis}
WASP-31 was observed in TESS Sector 9 (2/28/19-3/26/19) for 27 days with 2-minute cadence observations. To understand stellar variability, we used the out-of-transit Pre-search Data Conditioning Simple Aperture Photometry (PDCSAP) light curve. After masking the transits of WASP-31b and binning to $\sim$3.8 hours, the standard deviation of this light curve was 350 ppm. The strongest periodicity we identified using Lomb-Scargle periodigram analysis \citep{Lomb1976, Scargle1982} was that of a 2.88 
day period with a peak-to-trough amplitude of 510 ppm. Like the AIT analysis, this supports the idea that WASP-31 is a low activity---but not completely quiet---star. The period found with TESS is a third of that found with AIT. This type of aliased rotational period signal is common in the literature, \added{likely owing to the evolving topology of stellar activity regions and harmonic peaks due to insufficient data \cite[e.g.,][]{2016Gillon,2017Luger,2009Charbonneau,2011Berta,2018Mallonn}. Specifically, the AIT light curves could have too sparse sampling or the TESS monitoring could not be long enough.}

\subsubsection{ASAS-SN Analysis}
\begin{figure*}[htb]
    \centering
    \includegraphics[width=\textwidth]{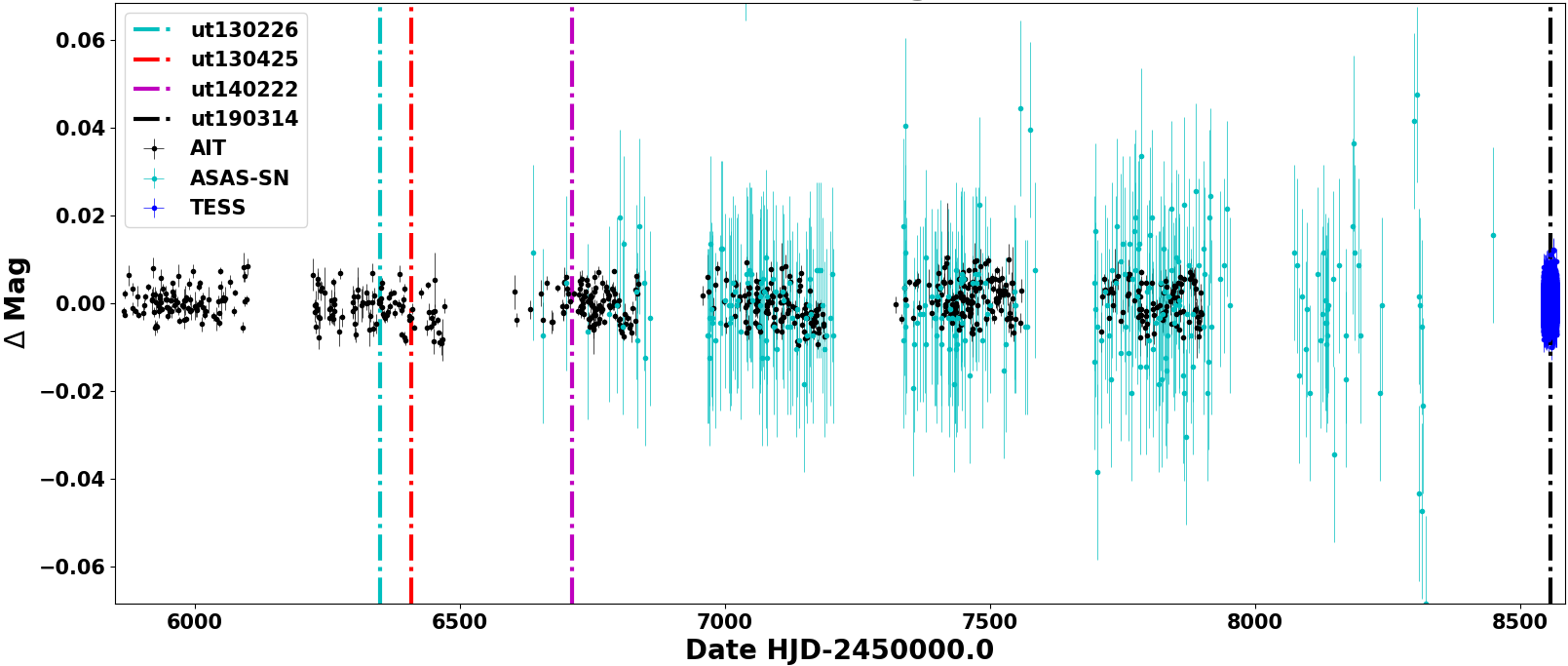}
    \caption{Photometric monitoring of WASP-31 with AIT (black), ASAS-SN (cyan), and TESS (blue). The times of the transits observed with Magellan are shown as dash dotted vertical lines. Given the differing observing bands of ASAS-SN (V), AIT (R), and TESS (650--1050\,nm), we limit our comparison to relative magnitude changes. Thus, the values plotted are the differences from the mean magnitude of each survey. All observations suggests that the star is relatively inactive, however, this cannot confidently be constrained with photometry alone.}
    \label{fig:PhotMon} 
\end{figure*}

\begin{figure*}[htb]
    \centering
    \includegraphics[width=\textwidth]{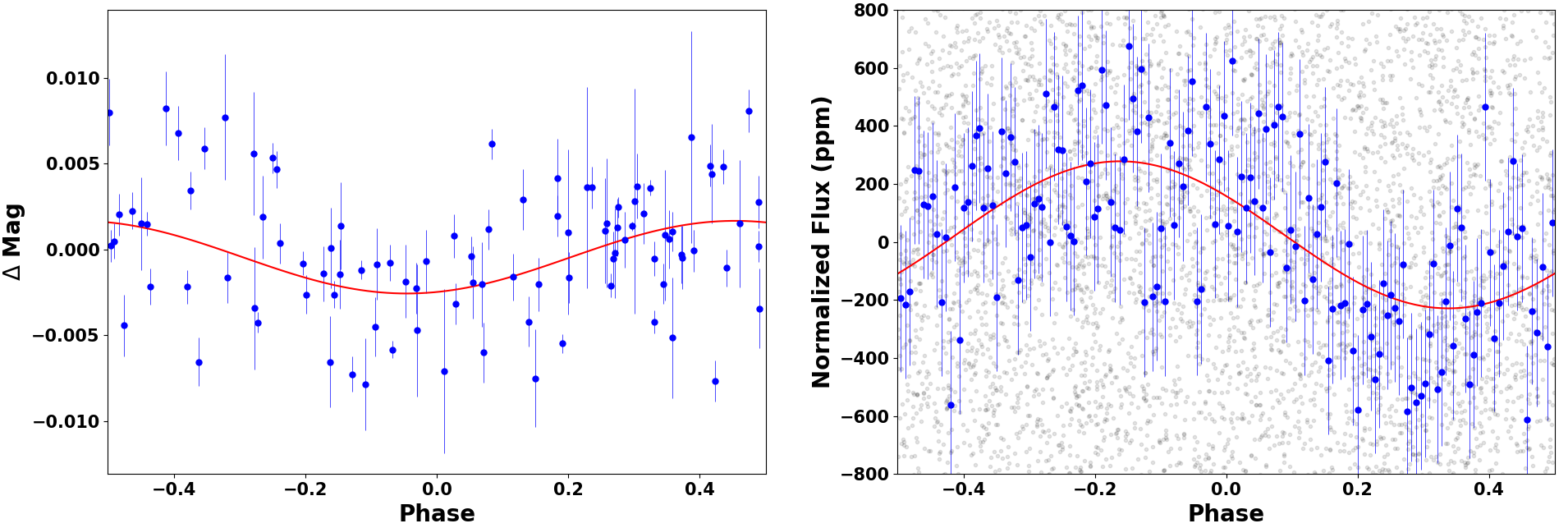}
    \caption{Phase curves of the AIT and TESS observations, showing low-amplitude periodicity in periodogram analyses. AIT data (left) show a 8.6-day periodicity with an amplitude of 0.004~mag. For the TESS data (right), the transparent gray points are all $\sim$15500 out-of-transit 2-minute cadence observations from TESS Sector~9. The blue points are the same data binned to portray the periodicity better. The red line is the 2.88~day, 507~ppm amplitude best-fit Lomb-Scargle model.}
    \label{fig:PhaseFPhotMon} 
\end{figure*}

ASAS-SN monitors the entire visible sky (V-band filter) with 24 telescopes all over the world. There are 262 observations of WASP-31 from December 2013 to November 2018 with two ASAS-SN cameras (bg, be) at the Cerro Tololo International Observatory, Chile and one camera (bc) at the Haleakala Observatory, Hawaii. The photometric precision of the ASAS-SN data is less than that of the AIT and TESS data, as seen in Figure~\ref{fig:PhotMon}. Thus, we cannot extract any evidence of stellar variability from the ASAS-SN light curves. Additionally, there is little ASAS-SN coverage of our transits; as such, we used the AIT and TESS data to constrain WASP-31's level of activity. The AIT and TESS phase-folded light curves are shown in Figure~\ref{fig:PhaseFPhotMon}.

Our conclusion from the analysis of the AIT, TESS and ASAS-SN datasets is that WASP-31b is a typical photometrically quiet F-type star. However, there is observational evidence that photometrically quiet F dwarfs are not completely inactive \citep[e.g,][]{2016Mercedes_K21b}. Furthermore, photometric monitoring alone is not enough to fully characterize stellar activity levels, as axisymmetrically distributed active regions do not contribute to rotational variability \citep{rackham2019}. Therefore, we still consider stellar activity in our retrieval analyses in Section \ref{sec:atmo_retriev}.

\section{Data Reduction \& Light Curve Analyses} \label{sec:data_red_lc_analysis} 
\subsection{Reduction Pipeline} \label{sec:data_red}
We reduced the raw IMACS data using the pipeline described in previous ACCESS papers \citep{Jordan:2013,rackham2017,Espinoza2019,Bixel2019,weaver2019access}. It performs standard bias and flat calibration, bad pixel and cosmic ray correction, sky subtraction, spectrum extraction, and wavelength calibration. The pipeline produces sets of extracted, wavelength-calibrated spectra for each epoch, for the target and each of the comparison stars in the field.

\added{The wavelength calibration was done by first fitting Lorentzian profiles to each spectral line on images taken with the calibration masks. Next, using these pixel positions and the known vacuum wavelengths, the wavelength solution for each spectrum was found by an iterative process in which a sixth order polynomial was fitted to the wavelengths as a function of pixel position, the data point with the greatest deviation from the fit was removed, and the process was repeated until the root mean square error value of the fit was less than 2 km s$^{-1}$ ($\sim$ 0.05{\AA}) \citep{rackham2017,Espinoza2017}.} 

The extracted, wavelength-calibrated spectra of WASP-31 for each of the four epochs are shown in Figure~\ref{fig:finExtSpec}, which highlights the gaps between CCDs. In the epochs observed with the f/4 camera, the spectra have three gaps. For the UT190314 transit, the one gap of the f/2 mode was close to the blue edge of the spectrum. This inhibited proper calibration of wavelengths shorter than $\sim$0.48 $\mu$m, which we excluded. 

The extracted spectra are then integrated across the entire wavelength range to produce white light curves or integrated over a narrow wavelength range to produce the binned light curves. These integrated white light curves are shown in the top panel of Figure \ref{fig:WLC}.

\subsection{Light Curve Detrending}\label{sec:lc_detrend} 
We first modeled the light curves following the approach of \cite{Jordan:2013}, using principal component analysis (PCA). For transit data (time-series measurements) we assume that the light curve $L_k(t)$ for a given comparison star $k$ is a linear combination of a set of signals $s_i(t)$, which represent the different instrumental and atmospheric effects on the light curves, i.e., 
\begin{align}
    L_k(t) = \sum_{i=1}^{K}A_{k,i}s_i(t),
\end{align}
where $A_{k,i}$ is the weight of each signal $s_i(t)$ in each light curve. 

With each light curve $L_k(t)$ as a row in a matrix $D$, we perform singular value decomposition\footnote{$Cov_t$[$s_i(t)s_j(t)$]=0 assuming that the signals are uncorrelated.}. This returns the eigenvectors $\vec{e}_i$ and eigenvalues $\lambda_i$ of the matrix. We obtain $s_i(t)$ from the product of the matrix of eigenvectors and the original data matrix $D$. Additionally, the importance of each signal for reconstructing $D$ is given by the eigenvalues $\lambda_i$ (largest $\lambda_i$, most important $\vec{e}_i$, etc.). This allows for the sequential fitting of those signals to our target star in order to find the simplest model that preserves the most important information present in the comparison stars. For further details of PCA, we refer the reader to \cite{PCAbookJolliffe2002}, or \cite{Espinoza2017} for its usage in high signal-to-noise time-series photometry. We implemented PCA with the same procedure as \cite{Jordan:2013} and \cite{rackham2017} and found 2, 4, 2, and 3 of the 4 PCA components to be the optimal number of components for each transit, chronologically\footnote{The optimal number of PCA components were determined with the k-fold cross-validation procedure \citep{2007Hastie}.}. 

\begin{figure*}[htb]
    \centering
    \includegraphics[width=\textwidth]{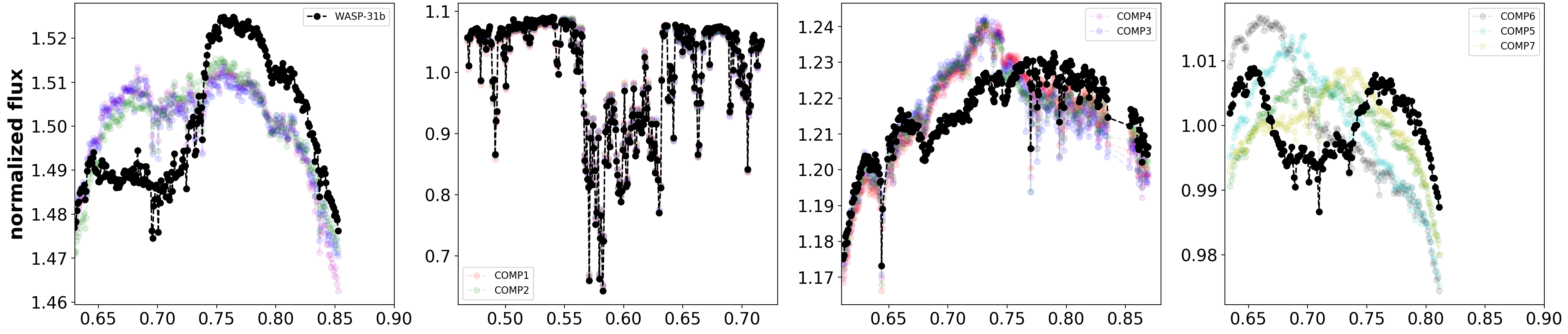}
    \includegraphics[width=\textwidth]{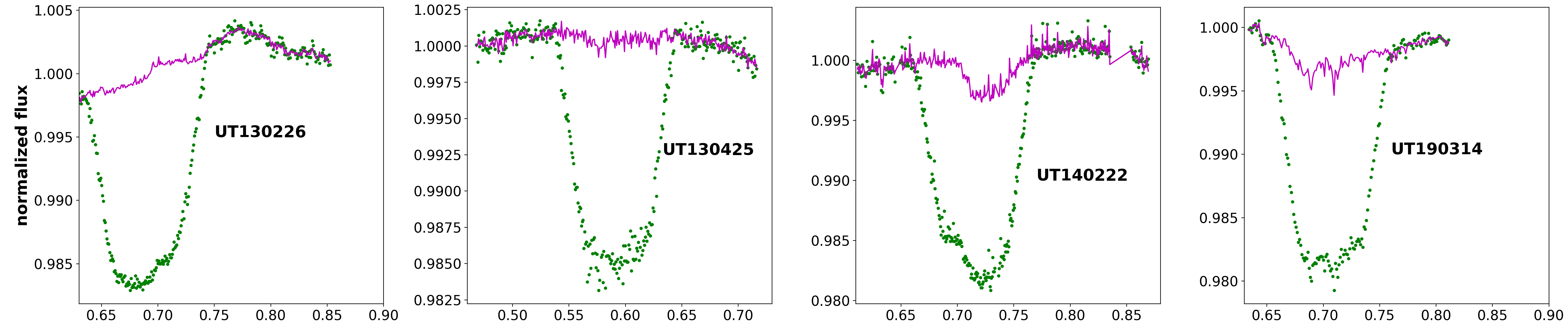}
    \includegraphics[width=\textwidth]{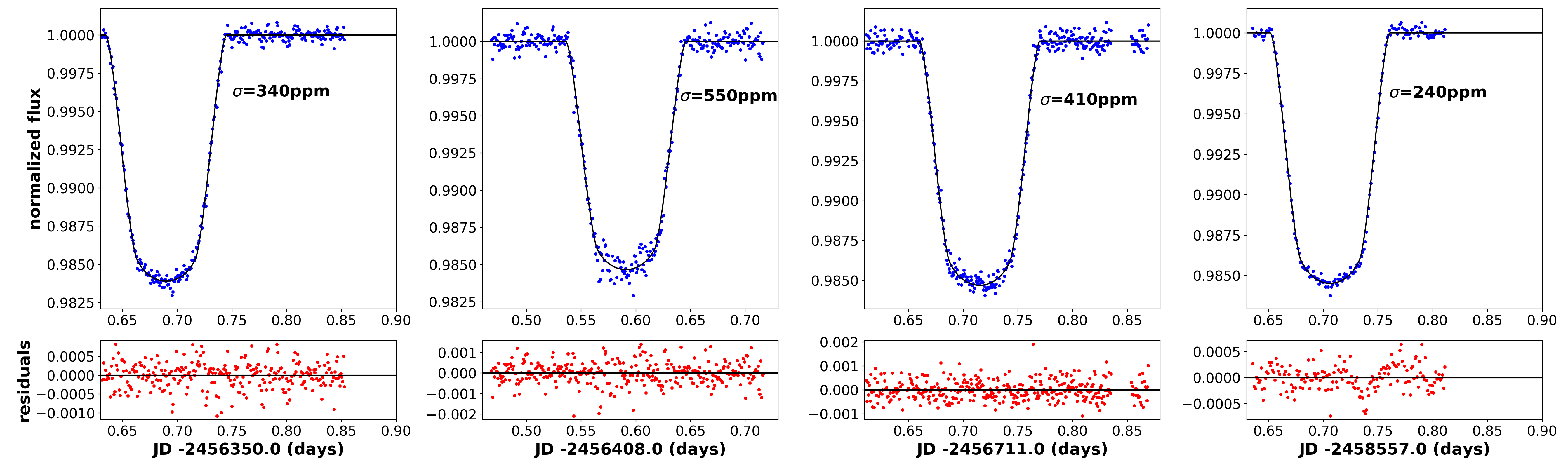}
    \caption{\textbf{Top:} Raw white light curves for our Magellan/IMACS observations directly from ACCESS's custom pipeline. Black is WASP-31 and transparent red, green, blue, magenta, cyan, grey, and yellow are comparison stars 1-7, respectively. See Table \ref{tab:comp_stars} for more information about the stars. \textbf{Middle:} The PCA-corrected light curves are shown in green. The magenta lines are GP fits on the light curves to correct for systematics using the 5 external parameters. This figure depicts that PCA correction alone is not sufficient to remove all systematics, and as such model averaging of PCA and GPs are justified. \textbf{Bottom:} Final white-light curve with the model-averaged PCA and GP systematics removed (blue points), along with the corresponding residuals (red points). The best-fitting transit light curve given our model averaging procedure is depicted as a solid black line. The value of $\sigma$ given in each panel is the standard deviation of the residuals in ppm. The middle panel is not an intermediate step to arrive at the bottom panel, given that the bottom panel is obtained directly when simultaneously doing PCA and GP model averaging of the data in the first panel. As can be seen in the right column, tranist UT130226 has little pre-transit data, which might cause systematics in its transmission spectra. This is explored further in Appendix \ref{appx:atmo_retriev_130226}.}
    \label{fig:WLC} 
\end{figure*}

However, as can be seen in the middle panel of Figure~\ref{fig:WLC}, using this PCA method alone was not sufficient to properly correct all systematics. This is because there are likely additional systematics that uniquely affect WASP-31 due to, e.g., color, detector, and instrumental differences with respect to the comparison stars. To account for this, we used Gaussian process (GP) regression to further model the target light curve in conjunction with PCA. A GP is a machine learning technique for non-parametric regression modeling \citep{Rasmussen_GP2005}, and has proven effective at detrending radial velocity data \citep[e.g,][]{2012Aigrain} and transmission spectroscopy \citep[e.g,][]{2012GibsonGPs}. \added{For our GP analysis, we define a joint probability distribution of the form $\mathcal{N}[0,\Sigma]$, where the covariance function ($\Sigma$) is defined as $K_{SE}(x_i,x_j) +\sigma^2_w\delta_{i,j}$. Here, $\sigma^2_w$, $\delta_{i,j}$, and $K_{SE}(x_i,x_j)$ are a jitter term, the Kroenecker delta function, and our \textit{kernel} respectively.}

\added{A multidimensional squared-exponential kernel was used for our GP regression: 
\begin{align}
  K_{SE}(x_i, x_j) = \sigma^2_{GP} \exp\Bigg(-\sum_{d=1}^{D} \alpha_d(x_{d,i}-x_{d,j})^2\Bigg),
\end{align}
where $\sigma^2_{GP}$ is the amplitude of the GP and $\alpha_d$ are the inverse (squared) length-scales of each of the components of the GP. The $x_i$ vectors here have components $x_{d,i}$, where each i denotes a time-stamp and where each d corresponds to a different external parameter.} The motivation for using the SE kernel is that it is 
a relatively smooth function and self-regulates the importance of the external parameters (\textbf{x}) \citep{Micchelli2006,Duvenaud2014}.

Because the joint probability of the data can be defined by a multivariate Gaussian using GPs, then the log marginal likelihood is known and given by:
\begin{align}
  \ln\mathcal{L} = -\frac{1}{2}r^T\Sigma^{-1}r - \frac{1}{2} \ln|\Sigma| - \frac{N}{2}\ln(2\pi).
\end{align}
Here $N$ is the number of measurements of the data. We used \texttt{george}\footnote{\url{https://github.com/dfm/george}} \citep{Mackey2014_george} to evaluate the marginalized likelihood of our dataset and \texttt{PyMultiNest}\footnote{\url{https://github.com/JohannesBuchner/PyMultiNest}} \citep{2014BuchnerPyMultiNest} (a python wrapper of \texttt{MultiNest} \citep{multinest2009}) to sample the hyperparameter space. In our analyses, we found including 5 external parameters $\textbf{x}$ useful to model the PCA corrected data with GPs. These time-dependent parameters were: i) the sky flux, ii) the variation of the FWHM of the spectra, iii) the airmass, iv) the position of the central pixel trace (perpendicular to the dispersion axis), and v) the drift of the wavelength solution.  

\added{The first 3 parameters are proxies of how Earth's atmosphere is affecting the stellar spectra. Over all exposures the FWHM ranges from 4.3 to 17.1 pixels and the airmass varied from 1.7 to 1.0 over the course of a night for each transit. The last two parameters are proxies for how slight physical variations in the instruments/detector could affect the spectra. Ultimately, exactly how these parameters affect the spectra is not well understood, but GPs are optimal at finding correlations between parameters and observables \citep{Rasmussen_GP2005}.} 

\added{The next task is to properly integrate the PCA, GP, and inherent transit model into one function defining our light curves. Our particular model used the magnitude of the target and comparison stars and was also used by \cite{Yan2020} to detrend ground-based spectroscopic transit data:
\begin{align}
  M_k(t) = c_k + \sum_{i=1}^{N_k} A_{k,i} s_i(t) - 2.51log_{10}T(t|\phi) + \epsilon,
 \label{equ:PCA+GP+transit}
\end{align}
where M$_k(t))$ is the (mean-subtracted) magnitude of the target star in the k$^{th}$ model, $c_k$ is a magnitude offset, $N_k$ is the number of PCA signals $s_i(t)$, $A_{i,k}$ is the weight for each signal in each models, $T(t|\phi)$ is the transit model with parameters $\phi$\footnote{We used the Python package \texttt{batman} \citep{Kreidberg2015_batman} to produce the analytic transit model.}, and $\epsilon$ is a stochastic component here modeled as a GP.}

Though PCA provides a weighting scheme for the importance level of the signals $s_i(t)$, the determination of which components improves our light curve model is still non-trivial. In order to incorporate our ignorance of the functional form of nuisance signals into our model fit, we used the model averaging technique outlined by \cite{Gibson2014_modelAvg}. \added{For our case, we averaged the posterior distributions of each M$_k(t))$, where the posteriors were explored using \texttt{PyMultiNest}. Our final model-averaged white light curves can be seen in the bottom row of Figure~\ref{fig:WLC}.}


\subsection{Light Curve Fitting}\label{sec:lc_fit} 
Using an improper limb-darkening (LD) law can introduce biases in the final light curve fit\citep{Espinoza2016_LDlaw}, and as such we used the open source package \texttt{ld-exosim}\footnote{\url{https://github.com/nespinoza/ld-exosim}} to determine the most appropriate LD law for our transit fits. The square root law was found to be best for all white light fits, given the data quality. \replaced{For the spectroscopic fits, described in Section~\ref{sec:trans_spec}, the logarithmic law was optimal for observation UT130226, but the square root law was best for the other observations.}{We used the same law for the spectroscopic fits, described in Section~\ref{sec:trans_spec}.} 

Next, we applied a $\sigma$-clipping to each of the white light curves to remove data points that deviated more than 3$\sigma$ from the model. This clipping was applied after obtaining an initial detrended light curve and corresponding best-fit model with all available data and lead to 3, 5, 0, and 1 data points being removed from epochs UT130226, UT130425, UT140222, and UT190314, respectively. 

We then used the 3-$\sigma$-clipped light curves to find the best transit parameters. The transit parameters used to fit the light curves were LD coefficients ($q_1, q_2$), planet orbital period ($P$), semi-major axis (in terms of star radius; $a/R_s$), the planet-to-star radius ratio ($p$), impact parameter ($b$), mid-transit time ($t_0$), inclination ($i$), eccentricity ($e$), and the argument of periapsis ($\omega$). We kept $e$ and $\omega$ fixed to 0 and 90$^{\circ}$ (assuming a circular orbit), following the results of \cite{Anderson2011_W31discov}. We placed Gaussian priors on all other parameters using the results of \citet{Anderson2011_W31discov}, then fit for them following the procedure described in Section \ref{sec:lc_detrend}. The best-fit white light curve orbital parameters for each epoch are summarized in Table \ref{tab:wlc_GP}.

\begin{deluxetable*}{CcRRRR}[htb]
    \caption{Fitted white light curve values. These are calculated using all integrated flux and available comparisons, unique to each epoch. This causes variation in depths per night, but improves the precision of the light curve parameters needed for binned light curve fits.}
    \label{tab:wlc_GP}
    \tablehead{\colhead{parameter} & \colhead{definition} & \colhead{UT130226} & \colhead{UT130425} &
    \colhead{UT140222} & \colhead{UT190314}}
    \startdata 
    p                & planet radius/star radius & 0.129\pm0.002    & 0.125\pm0.002             & 0.124^{+0.003}_{-0.006}    & 0.13\pm0.002    \\
    t_0-2.45e6      & mid-transit (JD) & 6350.6898^{+0.0004}_{-0.0005}       & 6408.5909\pm0.0002       & 6711.7150^{+0.0003}_{-0.0002} & 8557.7067\pm0.0002 \\
    P                & period (days)    & 3.405910\pm0.000005    & 3.405910\pm0.000005    & 3.405910\pm0.000005    & 3.405910\pm0.000005    \\
    $a/R_s$          & semi-major axis/star radius  & 8.15^{+0.10}_{-0.11}    & 8.04\pm0.09          & 8.13\pm0.09    & 8.15^{+0.07}_{-0.06}    \\
    b                & impact parameter & 0.772^{+0.007}_{-0.008}             & 0.784^{+0.007}_{-0.008}          & 0.769\pm0.006    & 0.775^{+0.006}_{-0.008}    \\
    i                & inclination (degrees)      & 84.6\pm0.1               & 84.4\pm0.1         & 84.6\pm0.1   & 84.5\pm0.1   \\
    q_1              & LD coeff 1       & 0.7\pm0.2                & 0.4^{+0.3}_{-0.2}          & 0.6\pm0.3    & 0.6^{+0.2}_{-0.3}    \\
    q_2              & LD coeff 2       & 0.4\pm0.2                & 0.5^{+0.2}_{-0.3}          & 0.7^{+0.2}_{-0.3}    & 0.8^{+0.1}_{-0.3}    
    \enddata 
\end{deluxetable*}

\section{Transmission Spectra}\label{sec:trans_spec}
To produce the optical transmission spectrum of WASP-31b, we generated transit light curves in the same manner as described in Section \ref{sec:data_red_lc_analysis}, but using a set of narrower wavelength bins. We propagated the flagged 3$\sigma$ outliers from the white-light analysis to each binned light curve. Our binning scheme was devised so it optimized spectro-photometric precision, while still letting us probe for atmospheric features, such as a scattering slope, and sodium and potassium lines. The average bin widths were 187\,{\AA}, where the lower-throughput, redder bins were made larger, and the regions around the Na\,I and K\,I features were 60\,{\AA} wide centered on each doublet (5892.9\,{\AA}, 7681.2\,{\AA}). 
We found 60\,{\AA} to be enough to encompass both doublet lines of each feature, providing high signal from potential features while minimizing suppression of signal from the surrounding continuum. This is also comparable to the 78\,{\AA} bin widths that \cite{Sing2015WASP31b} used to report a 4.2$\sigma$ K\,I detection. 

We employed the same analysis procedures as the white light curves (see Section \ref{sec:lc_fit}) but with $P$, $a/R_s$, $b$, $t_0$, and $i$ held fixed to values obtained from the corresponding white light curve fits, leaving only $q_1, q_2$, and $p$ as free parameters. The detrended light curves, their best-fit model, and residuals for each bin of each epoch are shown in Figures~\ref{fig:binnedLC1}--\ref{fig:binnedLC4} in Appendix \ref{Appx:Lgt_curves}. Tables \ref{tab:Trans_Spec_Each} and  \ref{tab:Trans_Spec_Combin} of Appendix \ref{Appx:Lgt_curves} list the values of $p$ obtained for each bin. 

\subsection{Combined Transmission Spectrum}\label{sec:combined_trans_spec}
The average precision of each bin for each spectrum is 0.007\,$R_p/R_s$. This is larger than the $\sim$0.006\,$R_p/R_s$ expected signal from Na and K features for this planet \citep{Sing2015WASP31b}. Therefore, we combined the transmission spectra from epochs UT130425, UT140222, and UT190314 to achieve higher precision. The UT130226 transit was not used in our combined data set because its transmission spectrum was inconsistent with the others, perhaps because of stellar activity or insufficient baseline before ingress. We discuss this further in Appendix~\ref{appx:atmo_retriev_130226}. 

\begin{figure*}[htb]
    \centering
    \includegraphics[width=\linewidth]{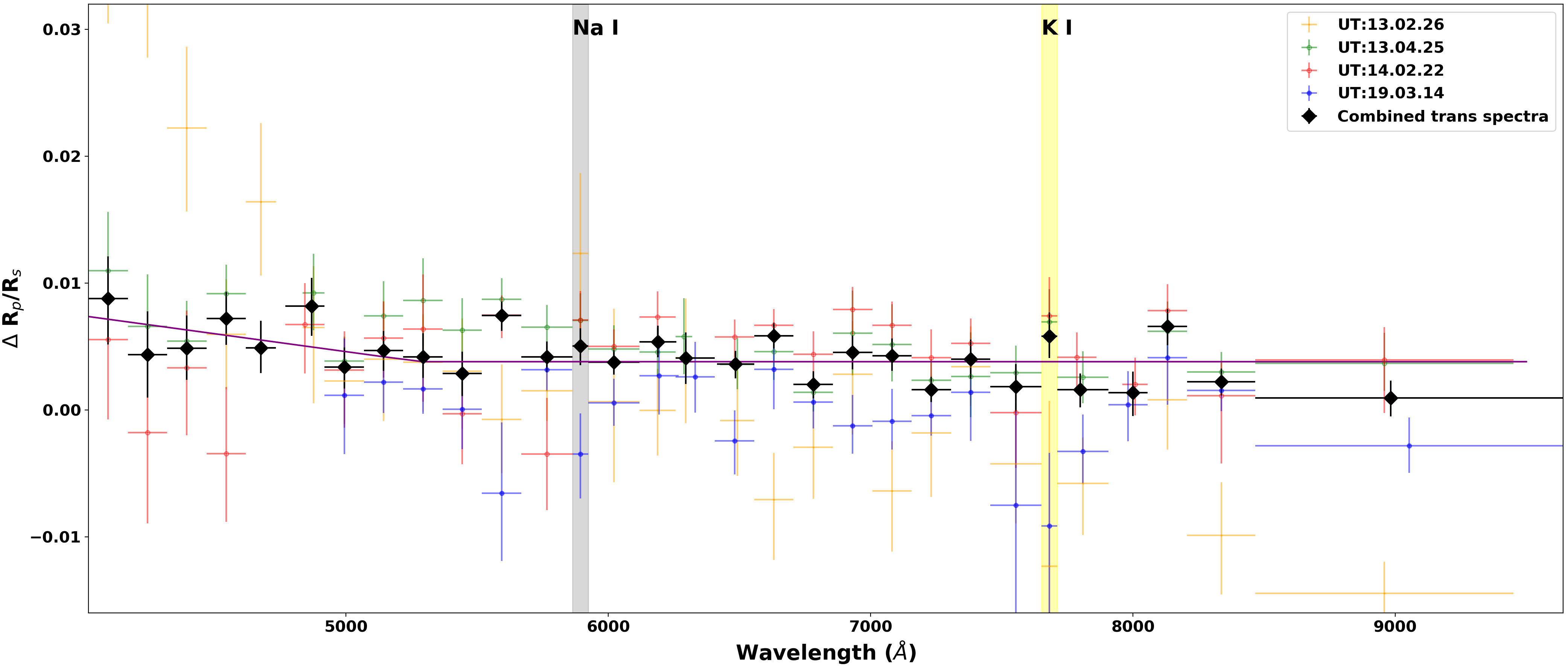}
    \caption{All four transmission spectra of WASP-31b taken by Magellan/IMACS, before an offset is applied, (faded colors) and the weighted average of the last 3 spectra (black). The first transit, UT130226 (transparent orange), significantly deviates from the other 3 spectra and has its highest $\Delta R_p/R_s$ of 0.0385, but cannot be seen in this figure because emphasis is placed on the other three consistent spectra. Transit UT130226 is likely an outlier because of insufficient baseline during observation or stellar activity (see Appendix \ref{appx:atmo_retriev_130226}). The purple line is a two part fit with a scattering slope bluewards of 5500{\AA} and a featureless spectrum redwards of 5500{\AA}, similar to fits done by \citep{Sing2015WASP31b} and \citep{Gibson2017WASP31}. The highlighted gray and yellow regions are centered around the Na I and K I lines, respectively.}
    \label{fig:MagTransSpec} 
\end{figure*}

Our spectra were combined by first applying an offset to each spectrum equal to the difference between their white light curve depths and the overall mean white light curve depth. We then combined the measurements for each spectroscopic bin by applying a weighted average using each point's $R_p/R_s$ errorbars as weights. The combined transmission spectrum, shown as black diamonds in Figure~\ref{fig:MagTransSpec}, has an average precision per bin of 0.0033\,$R_p/R_s$. We emphasize that when combining transmission spectra of different epochs, as we did, information about how the star and the terminator of the planet's atmosphere are behaving at a given epoch is lost, leaving only persistent (average) behaviors in the planet-star system. 

As can be seen in Table \ref{tab:wlc_GP}, the best-fit planet radius varies per epoch. These differences in our white light curve transit depths can be explained by the different locations of wavelength gaps in the detector for both the comparison stars and WASP-31, as shown in Figure~\ref{fig:finExtSpec}, which causes the integrated flux to be slightly different per epoch. Additionally, some epochs use different comparison stars as shown in Table \ref{tab:comp_stars}, which cause differences in white light curve flux ratios when dividing the flux of the targets by the flux of the comparison stars. We tested this hypothesis by using one matching comparison and a fixed wavelength range with no gaps for all datasets and found consistent transit depths for all three epochs. However, this approach significantly reduces the precision of the white light curves and the spectro-photometric bins, so we resorted back to applying offsets between epochs in order to use the full wavelength coverages and all comparisons. This approach is justified because the most important information from the transmission spectra is the relative changes in depth between wavelength bins. In addition, to address this issue, we fit for offsets between our combined transmission spectrum and other datasets in our atmospheric retrieval analysis, as described in Section \ref{sec:atmo_retriev}.

Figure~\ref{fig:MagTransSpec} shows the individual transmission spectra for each transit epoch, together with our combined transmission spectrum. Overlaid on the combined spectrum is a two-part fit with a scattering slope bluewards of 0.55~$\mu$m and a featureless spectrum redwards of 0.55~$\mu$m, similar to fits done by \citet{Sing2015WASP31b} and \citet{Gibson2017WASP31}. The fit is favored over a completely featureless spectrum ($\chi^2$ = 50.9) with a $\chi^2$ of 46.2.

\subsection{Previous Optical Transmission Spectra}\label{sec:previous_trans_spec}
\begin{figure*}[htb]
    \centering
    \includegraphics[width=\linewidth]{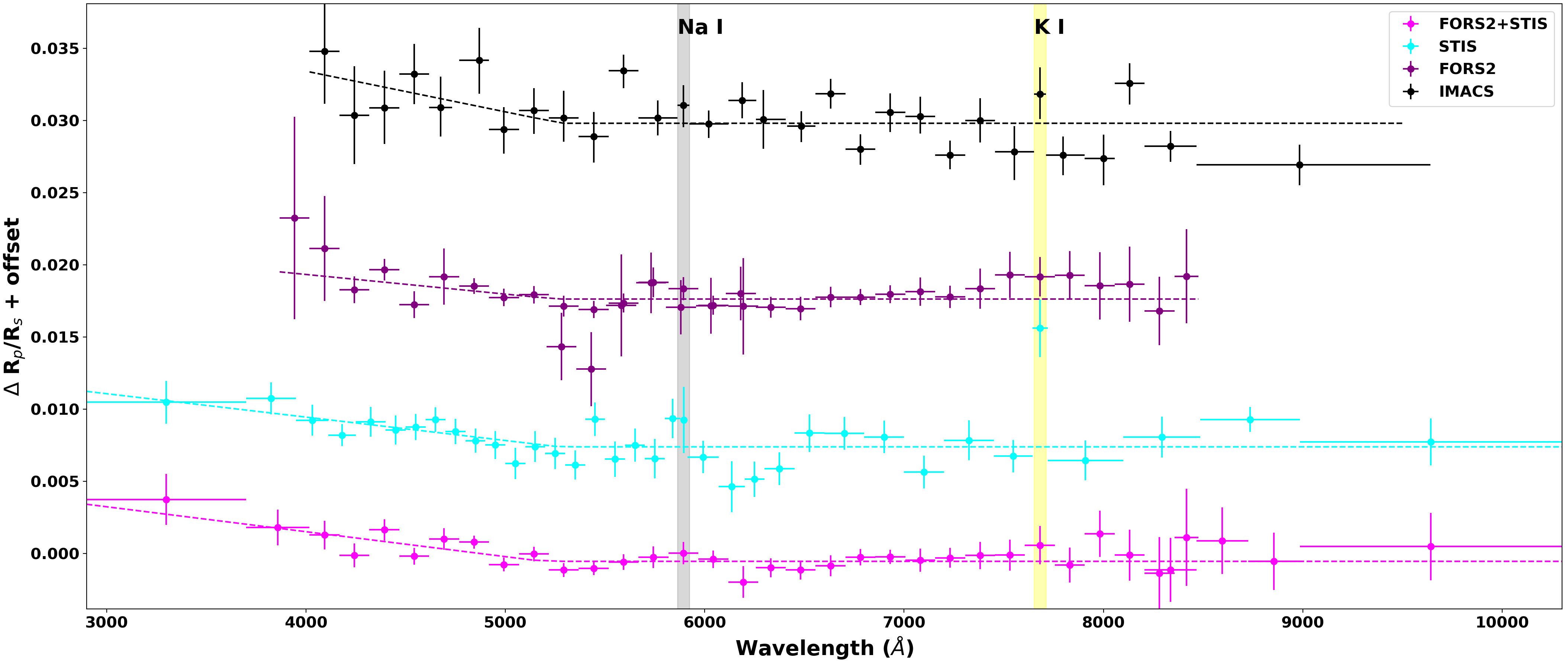}
    \caption{Transmission spectra of WASP-31b taken by Magellan/IMACS (excluding epoch UT130226, black), VLT/FORS2 \citep[purple, ][]{Gibson2017WASP31}, HST/STIS \citep[cyan, ][]{Sing2015WASP31b}, and combined FORS2 and reanalyzed STIS data \citep[magenta, ][]{Gibson2017WASP31}, with vertical offsets for clarity. Each spectrum has a two-part fit of a scattering slope blueward of 5500{\AA} and a featureless spectrum redward of 5500\,{\AA}, excluding the possible K signal centered at 7682.5\,{\AA} for the HST/STIS data, overlaid as dashed lines. All three spectra have signs of a blue scattering slope and featureless optical spectrum. The highlighted gray and yellow regions are centered around the Na I and K I lines, respectively.}
    \label{fig:AllTransSpec} 
\end{figure*}

All available low-resolution optical transmission spectra of WASP-31b are shown in Figure~\ref{fig:AllTransSpec}. When qualitatively comparing the Magellan/IMACS data with the VLT/FORS2 \citep{Gibson2017WASP31} and HST/STIS observations \citep{Sing2015WASP31b}, we see that all 3 spectra can be modeled with a scattering slope in the blue and a featureless spectrum beyond. \added{The simple models in Figure~\ref{fig:AllTransSpec} were created the same way as in Figure~\ref{fig:MagTransSpec}, just as a comparison with \cite{Sing2015WASP31b}'s model. However, in our retrieval analysis (Section~\ref{sec:atmo_retriev_Previous}), we do not use a two part model, but rather obtain scattering slopes with one continuous atmospheric model over the whole spectrum. This is also consistent with the findings of \cite{Barstow2016}, who found that a "split" spectrum with the HST data is not significantly favored compared to a continuous model.} Like the FORS2 observations, our IMACS observations show no signs of potassium or sodium absorption. This is contrary to the detection of potassium with HST/STIS. Though the overall shapes of the spectra agree with each other (aside from potassium absorption), we elected not to combine previous studies with our own in the subsequent analysis. This was because we are uncertain of the biases introduced to the data by the various detrending techniques for each analysis, and combining the spectra may lead to misinterpretation of signals lost or produced by compounded biases. This rationale is supported by the analysis of \cite{Gibson2017WASP31}, which found significantly different results for the same dataset depending on the detrending method. However, we did combine the IMACS optical transmission spectrum with near-IR HST/WFC3 spectra and mid-IR (\textit{Spitzer}/IRAC) data \citep{Sing2015WASP31b}, because there is only one observation/analysis for each and including them will help better characterize WASP-31b's atmospheric structure. In total, the combined transmission spectrum covers a wavelength range from 0.4$\mu$m to 5.06$\mu$m, with average bin sizes of 187{\AA} and 200{\AA} for the optical(Magellan/IMACS) and near-IR (HST/WFC3) data. 

\section{Atmospheric Retrieval Analyses} \label{sec:atmo_retriev}
We use atmospheric retrievals to provide a quantitative measurement of the existence of features in the transmission spectra. Given that results can depend on the priors and the specific retrievals used \citep[e.g.,][]{Kirk2019_retrievalComp}, we employed two independent retrievals to improve data interpretation: \texttt{PLATON}\footnote{\url{https://github.com/ideasrule/platon}} \citep{Zhang_2019PLATON}, and \texttt{Exoretrievals} \citep{Espinoza2019}.

\added{\texttt{PLATON} and \texttt{Exoretrievals} complement each other in several ways: \texttt{Exoretrievals} considers a small number of user defined molecules, but \texttt{PLATON} considers a large number (33) of pre-determined molecules. \texttt{Exoretrievals} performs free retrievals, allowing for abundances of individual molecules to be retrieved with non-equilibrium chemistry, whereas \texttt{PLATON} assumes equilibrium chemistry throughout the atmosphere. Thus, using both retrievals gives a more thorough understanding of the planet. Additionally, finding agreeing results in overlapping parameters, strengthens support of accurate interpretation of the data, given that they use very different fundamental assumptions.} 

In detail, \texttt{PLATON} uses the planet's limb temperature ($T$\textsubscript{p}), atmospheric metallicity (Z/Z$_\odot$), C/O ratio, cloud-top pressure ($P_0$), a scattering slope ($\alpha$), and an offset term between instruments to characterize observed transmission spectra using equilibrium chemistry. \added{\texttt{PLATON}'s sources of line lists for all 33 molecules considered are shown in Table 2 of \citep{Lupu2014}, which mostly, but not exclusively, comes from HITRAN. \texttt{PLATON} includes collision induced absorption, where all coefficients come from HITRAN.} \texttt{PLATON} also simultaneously fits for stellar activity using the occulted temperature of the star ($T$\textsubscript{star}), the temperature of unocculted spots ($T$\textsubscript{spot}), and their area ($f_\mathrm{spot}$). \texttt{PLATON} uses the nested sampling algorithm \texttt{dynesty}\footnote{\url{https://github.com/joshspeagle/dynesty}} \citep{Speagle_2019Dynesty} to explore the parameter space.
 
\texttt{Exoretrievals} uses a semi-analytical formalism to model the planet with an isothermal, isobaric, atmosphere with non-equilibrium chemistry \citep[see][]{2017Heng}. \added{The cross sections we considered with \texttt{Exoretrievals} were CO, CO2, CH4, TiO, NH3,  FeH, HCN, H2O, Na, and K where the line list of the molecules came from HITEMP \citep{Rothman2010}, ExoMol \citep{Yurchenko2013, Tennyson2016}, and TiO was synthesized using the latest (2012 version) line lists calculated by B. Plez, obtained from VALD \citep{Ryabchikova2015}). The Na and K cross-sections were determined from analytical Lorentzian-profiled doublet shapes used in \cite{MacDonald2017}. Unlike \texttt{PLATON}, \texttt{Exoretrievals} does not include collision induced absorption.} However, it does also account for stellar photospheric heterogenities by modeling the occulted and unocculted regions of the stellar photosphere with PHOENIX models \citep{2013Husser}, following the approach of \cite{Rackham2018,rackham2019}. The retrieval has three parameters to characterize the stellar surface: the temperature of the region occulted by the planet ($T$\textsubscript{occ}), the temperature of unocculted heterogeneities ($T$\textsubscript{het}), which can be hotter or colder than the mean photosphere, and the heterogeneity covering fraction ($f_{het}$). The planet's atmosphere is characterized by a factor ($f$) multiplied to the planetary radius to find a reference radius at which the atmosphere is optically thick (i.e. $R_0 = fR_p$), the corresponding pressure where the atmosphere is optically thick ($P_0$), the limb atmospheric temperature (T\textsubscript{p}), mixing ratios of specified species (i.e. Na, K, and H$_2$O), offset terms for different instruments, and two parameters to characterize the cloud/haze extinction ($\kappa$). The two cloud/haze parameters are a \textit{Rayleigh-enhancement factor} ($a$) and a \textit{cloud/haze power law} ($\gamma$), and they quantify extinction with:
\begin{align}
    \kappa(\lambda) = a\sigma_0(\lambda/\lambda_0)^{\gamma},
\end{align} 
where $\sigma_0 = 5.31\times10^{-31} \mathrm{m}^2$ and $\lambda_0 = 350\mathrm{nm}$ \citep{2017_Sedaghati,2017MacDonald}. The code uses nested sampling with \texttt{PyMultiNest} to measure the evidence for each of the retrieved models. The priors we used for both retrievals are shown in Table \ref{tab:Priors}. Because \texttt{Exoretrievals} allows for individual molecules to be tested, we used the results of \texttt{Exoretrievals} to determine the presence of water, potassium, or sodium.

\begin{deluxetable*}{|C|C|C|C|C|C|}[htb]
    \caption{The priors for \texttt{Exoretrievals} and \texttt{PLATON}. These priors were set to allow for a wide parameter space to be surveyed, but contained within physical regimes. Not all parameters were included in each model fit (see Tables \ref{tab:lnZAllMag}). We used 10000 live points for each \texttt{Exoretrievals} run and 5000 for \texttt{PLATON}. For further description of the parameters of \texttt{Exoretrievals}, please refer to the Appendix of \citet{Espinoza2019}.}
    \label{tab:Priors}
    \tablehead{\multicolumn{3}{|c|}{\bfseries {\large Exoretrievals}} &\multicolumn{3}{|c|}{\bfseries {\large PLATON}}}
    \startdata 
    \textbf{parameter}                          & \textbf{prior function} & \textbf{prior bounds} & \textbf{parameter}           & \textbf{prior function} & \textbf{prior bounds} \\ \hline
    \text{cloud top}                            & \text{log-uniform}      & \text{-8 to 2}        &  \text{cloud top }           & \text{log-uniform} & \text{-3 to 7} \\
    \text{ pressure (P\textsubscript{0}, bars)} &                         &                       &  \text{pressure (P\textsubscript{0}, pascals)}        &                    &  \\ \hline
    \text{planetary atmospheric}                & \text{uniform}          & \text{800 to}   & \text{planetary atmospheric} & \text{uniform} & \text{800 to} \\
    \text{temperature (T\textsubscript{p})}                      &                         &  \text{1900K}                     & \text{temperature (T\textsubscript{p})}        &                &  \text{1900K}\\ \hline
    \text{stellar temperature)}& \text{uniform} & \text{6000.0 to}& \text{stellar temperature }& \text{gaussian}         & \text{mean=6300K,} \\ 
    \text{(T\textsubscript{occ})}   &      &  \text{6300.0K}               & \text{(T\textsubscript{star})}   &  & \text{std=150K} \\ \hline
    \text{stellar heterogeneities}              & \text{uniform}          & \text{2300.0 to}& \text{stellar heterogeneities} & \text{uniform} & \text{5900.0 to} \\
    \text{temperature (T\textsubscript{het})}   &                         & \text{7000.0K }& \text{temperature (T\textsubscript{spot})} &    &  \text{6700.0K} \\\hline
    \text{spot covering}                        & \text{uniform}          & \text{0 to .8}          & \text{spot covering}                                      & \text{uniform} &\text{ 0 to .8} \\ 
    \text{fraction (f\textsubscript{het})}      &                         &               & \text{fraction (f\textsubscript{spot})}                                      &                   &                   \\ \hline
    \text{offset (depth)}                       & \text{gaussian}         & \text{mean=0,} & \text{offset (depth)}                   & \text{uniform} & \text{-8400 to} \\ 
                                                &                         & \text{ std=1000ppm} &                   &  &  \text{8400ppm}\\ \hline
    \text{cloud/haze amplitude ($a$)}                 & \text{uniform }         & \text{-30 to 30}           & \text{scattering factor}                                           & \text{log-uniform} & \text{-10 to 10}\\ \hline
    \text{cloud/haze power law (}$\gamma$\text{)}\tablenotemark{a}               & \text{uniform }         & \text{-10 to 4}            & \text{scattering slope (}$\alpha$\text{)}\tablenotemark{b} & \text{uniform} & \text{-4 to 10}\\ \hline
    \text{cloud absorbing} & \text{log-uniform} & \text{-80 to 80}        & \text{metallicity (Z/Z\textsubscript{odot})}     & \text{log-uniform} & \text{-1 to 3}\\
    \text{cross-section (}$\sigma$\text{\textsubscript{cloud})} &                   &      &                                            &  & \\ \hline
    \text{trace molecules'}       & \text{log-uniform}      & \text{-30 to 0 }             & \text{C/O}                                                         & \text{uniform} & \text{0.05 to 2}\\
    \text{mixing ratios}       &     &            &                                                        &  & \\ \hline
    \text{reference radius factor} ($f$)          & \text{uniform }         & \text{0.8 to 1.2}            & \text{reference radius (R\textsubscript{p})}                       & \text{uniform} & \text{1.35 to 1.6R\textsubscript{j}}\\ \hline
    \enddata
\tablenotetext{a}{This is the exponent of the scattering slope power law, where $-4$ is a Rayleigh scattering slope.}
\tablenotetext{b}{This is the wavelength dependence of scattering, with 4 being Rayleigh.}
\end{deluxetable*}

\subsection{Combined Magellan Data} \label{sec:atmo_retriev_CombMag}
We run \texttt{PLATON} and \texttt{Exoretrievals} on the combined IMACS, HST/WFC3, and \textit{Spitzer} data. \added{For each retrieval run, we fit for an offset between the optical data and the IR data but no offset between the \textit{Spitzer} and the HST/WFC3. The \textit{Spitzer} and HST/WFC3 data was taken as a single IR dataset, where \cite{Sing2015WASP31b} analyzed them both using the same ephemeris and system parameters.} Using the retrievals we tested for different scenarios: 1) a featureless exoplanet atmosphere, where all potential signals are covered by high-altitude clouds, 2) a spectrum with any signal detected in the spectrum coming from active regions on the surface of the star, 3) an atmosphere with scatterers, and 4) an atmosphere with scatterers, and stellar activity contamination. Additionally, we tested \texttt{Exoretrievals}' models with the same 4 scenarios, but including only H\textsubscript{2}O, Na, or K, individually and simultaneously in order to isolate their atmospheric effects. 

Table \ref{tab:lnZAllMag} summarizes the Bayesian evidences of each scenario with respect to a featureless spectrum ($\Delta \ln Z$). Following \cite{2008Trotta} and \cite{2013Benneke}, the frequentist significance of those $\Delta \ln Z$ values scales as: |$\Delta \ln Z$| of 0 to 2.5 is inconclusive with < 2.7$\sigma$ support for the higher evidence model, |$\Delta \ln Z$| of 2.5 to 5 corresponds to a moderately significant detection of 2.7$\sigma$ to 3.6$\sigma$, and |$\Delta \ln Z$| $\geq$ 5 corresponds to strong support for one model over the other.

\added{\texttt{Exoretrievals} could also fit for CO, CO\textsubscript{2}, CH\textsubscript{4}, TiO, NH\textsubscript{3}, FeH, and HCN. Given the temperature\citep{Fortney2008,Evans2018}, wavelength range, and likely high altitude clouds (see discussion in section \ref{sec:retriev_interp}) we do not expect to detect any of those 7 species. As a precaution, we did run \texttt{Exoretrievals} with the 3 cases discussed above (featureless, activity, scatters) and including each of the 7 species individually. As expected, there was insufficient evidence to support any of those models.}

The best model fit from \texttt{Exoretrievals} was a flat spectrum. However, the |$\Delta \ln Z$| between the featureless model and the majority of the others was below 2.5, implying that those models cannot be distinguished from a flat spectrum. For \texttt{PLATON} the best fit was a model with a near-Rayleigh scattering slope (5.3$^{+2.9}_{-3.1}$), log metallicity of 1.3$^{+1.1}_{-1.4}$, and  C/O = 0.52$^{+0.87}_{-0.32}$, but again the support for this model is indistinguishable from the others ($\Delta \ln Z$ = 1.7). The undefined features (i.e lack of Na or K) and nominal detection of a scattering slope found here is consistent with the results of \cite{Gibson2017WASP31}. \citeauthor{Gibson2017WASP31} found no significant detection of K or Na and had only a tentative detection of a scattering slope with their FORS2 data. It was only when including the bluer HST/STIS data between 0.3 and 1.0 $\mu$m that they confidently detect a scattering slope. Unfortunately our Magellan data does not reach that far into the blue.

\added{Though the highest model with \texttt{Exoretrievals} was that of a featureless atmosphere, in order to compare results from both retrievals, we examine the \texttt{Exoretrievals} scattering model with the highest evidence ($\Delta \ln Z$ = -1.77). This model is still indistinguishable from the featureless model.} We find that all overlapping parameters agree with each other within their uncertainties. The planet's terminator temperature of T\textsubscript{p} = 1223$^{+429}_{-292}$ K with \texttt{Exoretrievals} agrees with the calculated equilibrium temperature \citep[1575K][]{Anderson2011_W31discov}, and the T\textsubscript{p} of 1219$^{+262}_{-245}$ K retrieved with \texttt{PLATON} is nearly 1-$\sigma$ away from the calculated equilibrium temperature.
A low retrieved terminator temperature is not uncommon, and an inherent bias produced when using 1-D transmission spectrum models for a 2-D problem \citep{Pluriel2020,McDonald2020}. However, using equation 1 of \cite{2007Mercedes} to calculate the equilibrium temperature of the planet, we find the retrieved temperatures are still consistent if the planet has a bond albedo above 0.3\footnote{assuming efficient heat distribution, $f$ =1/4, and using the retrieved system parameters}. This might be a slightly high albedo for a hot Jupiter but is completely reasonable relative to the solar system's jovians \citep{2008Rowe,Mallonn2019}. 

The log cloud top pressure of -2.9$^{+3.3}_{-3.2}$ bars with \texttt{Exoretrievals} and -3.6$^{+2.7}_{-2.1}$ bars with \texttt{PLATON}, while not well constrained, implies that there are high-altitude clouds. High-altitude clouds provide a good explanation for the data, given that features such as Na and K are not present, the possible water feature in the IR data is strongly muted, and the retrieved parameters have large uncertainties. But again, it cannot be confidently claimed that this is the case because the lack of convergence of the cloud top pressure. The $\gamma$ of -3.8$^{+3.1}_{-3.8}$ obtained with \texttt{Exoretrievals} and scattering slope of 5.3$^{+2.9}_{-3.1}$ with \texttt{PLATON} are within 1$\sigma$ of a Rayleigh scattering slope, which was detected at some level by previous studies \citep{Sing2015WASP31b, Gibson2017WASP31}.

\begin{figure*}[h!]
    \centering
    \includegraphics[height=1\columnwidth]{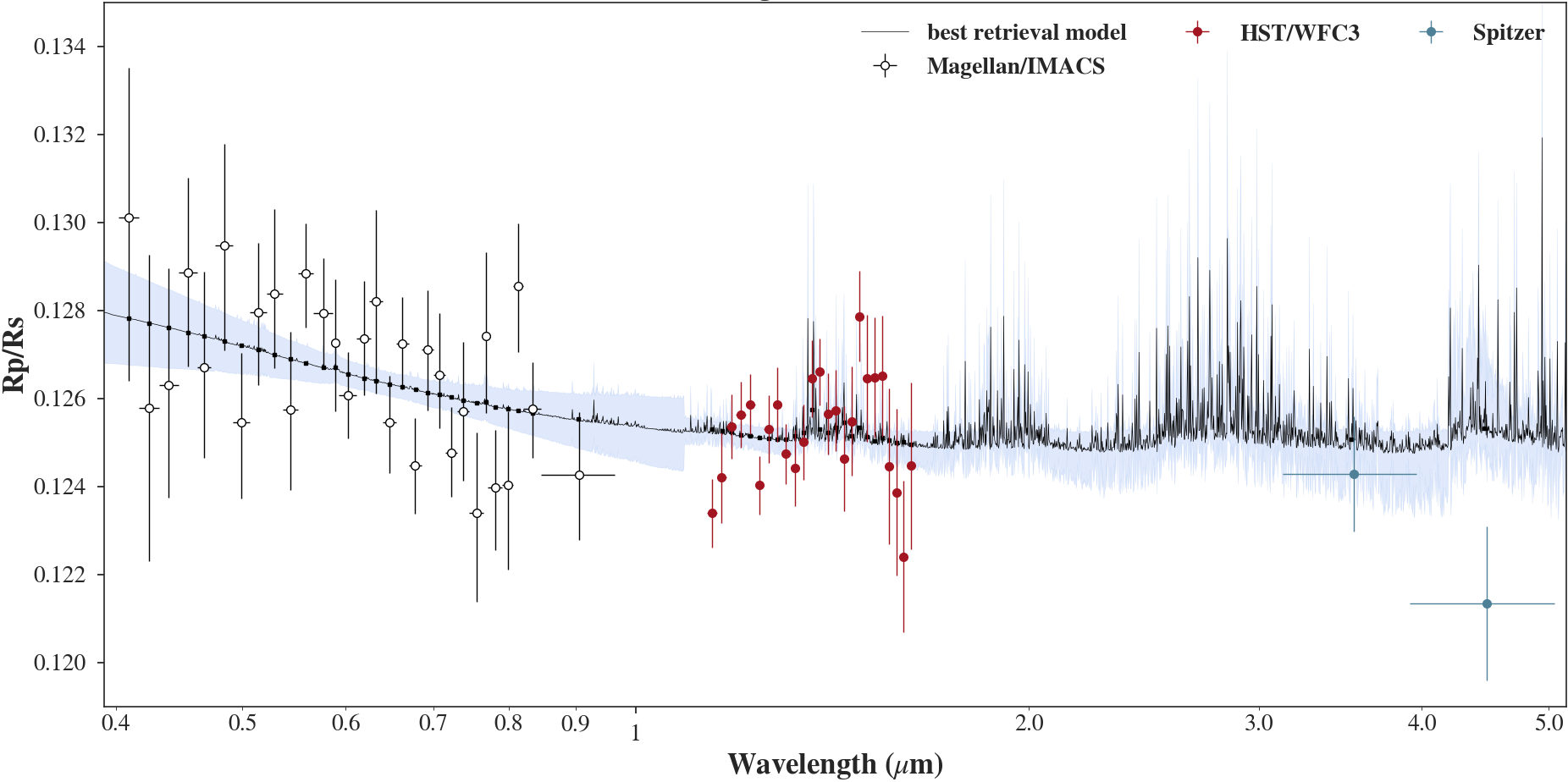}
    \caption{Combined transmission spectrum of the three usable Magellan/IMACS optical transits, HST/WFC3, and \textit{Spitzer} IR transits. With this data the best-fit model (highest $\Delta\ln Z$) was one utilizing \texttt{Platon} and obtained a scattering slope consistent with Rayleigh (5.3$^{+2.9}_{-3.1}$), log metallicity (Z/Z$_\odot$) of 1.3$^{+1.1}_{-1.4}$, C/O ratio of 0.52$^{+0.87}_{-0.32}$, planet limb temperature of 1220$^{+260}_{-250}$ K, and log cloud top pressure of -3.6$^{+2.7}_{-2.1}$ bars. The solid black line is the retrieval's best fit (highest evidence), the shaded region is the 1-$\sigma$ confidence interval, and the colored dots with error bars are the data for various instruments.}
    \label{fig:PlatIMACS_TranSpec} 
\end{figure*}

The highest $\Delta \ln Z$ transmission spectrum model retrieved is that of a scattering slope with \texttt{PLATON}, and is overlayed on top of our combined Magellan/IMACS data in Figure~\ref{fig:PlatIMACS_TranSpec}, with corner plot results shown in Figure~\ref{fig:PlatIMACS_Corn}.

\added{Previous detailed atmospheric studies on WASP-31b's atmosphere with data introduced in \cite{Sing2015WASP31b} are consistent with our retrieval analysis using IMACS data introduced in this manuscript. Specifically, \cite{wakeford2017} fit models to the HST/STIS WASP-31b data and found that clouds made of enstatite (MgSiO\textsubscript{3}) and iron act as a gray opacity source from log pressures of -4.1 to -1.7 in WASP-31. This agrees with the cloud deck pressure, and muted features we retrieved. In fact, given the high uncertainties of our retrieved scattering slope, a near gray (little color dependent scattering) cloud deck is also supported. \cite{Barstow2016} fit 3600 atmospheric models to the HST/STIS WASP-31 data, and found a best fit for a gray cloud deck at 100mbars, but there were a few high evidence models with a Rayleigh slope and cloud top pressure as low as 0.01mbars. \cite{Pinhas2017} also found multiple indistinguishable fits for the HST/STIS WASP-31b data with their models, but had a best fit slope of -5.52+/-1.27 which agrees with our best fits. These findings in conjunction with ours emphasizes the difficulty in converging on the exact picture of WASP-31b's atmosphere, due to cloud formation.}

\begin{deluxetable*}{|l|C|C|C|C|C|C|C|C|}[h!]
    \caption{$\Delta\ln Z$ for various \texttt{Exoretrievals} (left) and \texttt{PLATON} (right) models relative to a featureless spectrum with the subset of data that included the combined Magellan/IMACS (excluding transit UT130226), HST/WFC3, and \textit{Spitzer} IR data. For \texttt{Exoretrievals} no model is significantly preferred, but a flat spectrum has the highest evidence. For \texttt{PLATON} a spectrum with an optical scattering slope is slightly favored, but indistinguishable from the other models. }
    \label{tab:lnZAllMag}
    \tablehead{\multicolumn{6}{|c|}{\bfseries {\large Exoretrievals}} &&\multicolumn{2}{|c|}{\bfseries {\large PLATON}}}
    \startdata 
      \textbf{Model:}  &  \text{featureless}  &  $H_2O$  &  $Na$  &  $K$  &  $H_2O + K +Na$  && \textbf{Model:} &\\ \hline
      featureless &                0.0 &	-0.54 &	-1.09 &	-0.99 &	-1.19&& \text{featureless} & 0.0\\
      scatterers &         --- &	-1.77 &	-2.19 &	-2.42 &	-2.48& &\text{scattering} & 1.7\\
      activity &            -0.29 &	-0.41 &	-0.79 &	-0.97 &	-1.1& &\text{stellar activity} & 0.63\\
      scatterers \& activity & --- &	-2.66 &	-3.09 &	-2.72 &	-3.35 & &\text{Both} & 1.04
    \enddata
\end{deluxetable*}

\subsection{Previous Data} \label{sec:atmo_retriev_Previous}
In order to test the consistency of our IMACS observations and previous observations of WASP-31b, we ran the same retrieval analysis against the three separate optical data in the literature, which were the original HST/STIS \citep{Sing2015WASP31b}, VLT/FORS2, and the combined FORS2 and reanalyzed STIS data \citep{Gibson2017WASP31}. We combined each of these subsets of data with the WFC3 and \textit{Spitzer} IR data. 

In summary, the VLT/FORS2 retrieval analysis produced nearly identical results to our IMACS analysis. That is,   inconclusive support for a flat spectrum with \texttt{Exoretrievals} and a slightly favored, indistinguishable from a flat line ($\Delta\ln Z$ = 0.26), near-Rayleigh optical scattering slope with \texttt{Platon}. Given the similar wavelength coverage of IMACS (0.4--0.96$\mu$m) and FORS2 (0.39--0.85 $\mu$m) and their precision (0.0033$R_p/R_s$), we expect to see this agreement.

We detected potassium ($\Delta \ln Z$=3.8) and a scattering slope with \texttt{Exoretrievals} when performing retrieval analysis with the HST/STIS data. Interestingly, the best model with \texttt{Platon} was one with strong support ($\Delta \ln Z$=20.9) towards only stellar activity, which is likely a portrayal of a degeneracy between scattering slope and stellar activity \citep{rackham2017}.

The evidences obtained with \texttt{Exoretrievals}, utilizing the combined FORS2 and reanalyzed STIS data subset, favored the model with water and a scattering slope at $\Delta \ln Z$=2.5. This data subset also moderately favored a scattering slope with \texttt{Platon} at $\Delta \ln Z$=4.2. 

With both the STIS and combined FORS2 and reanalyzed STIS data subsets, the retrieved scattering slopes were higher (> 2$\sigma$) than Rayleigh. This is likely due to the fact that these data subsets include bluer (0.3--0.4$\mu$m) wavelengths, which constrain the high scattering slope. We tested this by running the combined FORS2 and reanalyzed STIS data excluding the first two datapoints (centered at 3350 and 3859{\AA}) against \texttt{Platon}\footnote{\texttt{Platon} is only capable of calculating the spectrum from 0.3 to 30.0$\mu$m. Therefore, we had to shift the central wavelength of the bluest datapoint by 50{\AA}. To do so, we did a linear interpolation between the three bluest bins, using the \texttt{linear} method of the \texttt{scipy.interpolate} package \citep{Virtanen2020}. We found a 45ppm change in depth.}, and obtained a shallower slope that was consistent with Rayleigh scattering at 1$\sigma$.

\section{Retrieval Interpretation} \label{sec:retriev_interp}

\subsection{Optical Scattering Slope} \label{sec:opt_scat_slope}
As discussed in Sections \ref{sec:atmo_retriev_CombMag} and \ref{sec:atmo_retriev_Previous}, the Bayesian evidences of the FORS2 or IMACS datasets were unable to distinguish a flat spectrum from that with a scattering slope. The models that fit for a scattering slope with these data did retrieve slopes that were within Rayleigh, but those retrieved slopes had large uncertainties. In contrast, retrieval runs utilizing the STIS or combined FORS2 and reanalyzed STIS data subsets moderately detected a scattering slope. These slopes were above Rayleigh (|$\gamma$| = $\alpha$ = 8.3) and better constrained with approximate error bars of $\pm$ 1.5. However, the STIS dataset is also the only one where K and stellar activity are significantly retrieved, portraying our overall uncertainty of its retrieved parameters. Thus, it is unclear whether the slope anchored by the bluer STIS datapoints implies that WASP-31b does have a large scattering slope or that the STIS dataset is an outlier. As such, we can only report that there are tentative signs of an optical Rayleigh scattering slope redward of 0.4$\mu$m.

\subsection{Sodium \& Potassium Features} \label{sec:Na+K_features}
Of the four available WASP-31b optical datasets, only the STIS data had a Bayesian evidence that favored potassium absorption. As \cite{Gibson2017WASP31} imply, this is likely produced as a side effect of the detrending technique used. Another cause may be that the star was active during the HST observations, which \cite{Pont:2013}, \cite{rackham2017}, and \cite{Espinoza2019} have shown could mimic transmission spectral signals. As introduced in Section \ref{sec:atmo_retriev_Previous}, the Bayesian evidence using \texttt{Platon} strongly favored a model characterizing the STIS transmission spectrum with stellar activity. This is supported by the the AIT photometric monitoring that only showed signs of activity during the 2012-2013 season, which coincides with the STIS observations taken in June of 2012. However, Keck/HIRES observations of WASP-31's Ca II H \& K emission lines suggest that the star was extremely quiet (logR'\textsubscript{HK}= -5.225) when observed (2002-2009) \citep{Sing:2016}. Thus, it is possible, while not certain, that stellar activity could contribute to the cause of this unique transmission spectrum. No matter the cause of the potassium signal in the original STIS data, given that there are two independent low-resolution transmission spectra and a high-resolution transmission spectrum that show no signs of K, we conclude that WASP-31b shows no optical atomic absorption. 

It could also be argued that because the K feature (7681{\AA}) is near prominent tellurics ($\sim$7580--7680{\AA}), ground-based observations would not be able to detect potassium. However, there are multiple cases where K is detected from the ground \citep{Sing2011,Knicole2012,Chen2018} with low resolution R$\sim$500, and high resolution can deconvolve Earth's spectroscopic signal from a target's \citep{Gibson2019WASP31}. Thus, the Magellan/IMACS observations further paints the picture of WASP-31b lacking alkali absorption features. This is indicative of high-altitude clouds present in the planet's atmosphere, given that prominent Na and K absorption is expected in hot Jupiters like WASP-31b \citep{Seager:2000}. \added{Where authors such as \cite{Morley2012} and \cite{Gao2020} have found that for gas giants at the temperature range of WASP-31b, silicate clouds can heavily alter observed spectral features.}

\subsection{Water Features} \label{sec:H2O}
None of the fits in Sections \ref{sec:atmo_retriev_CombMag} and \ref{sec:atmo_retriev_Previous}, aside from the FORS2 and reanalyzed STIS data subset, showed significant detection of H\textsubscript{2}O. However, in general the models that include H\textsubscript{2}O are more favored (see Table \ref{tab:lnZAllMag}), suggesting that there are signs of water features in the IR data. This is consistent with the findings of \cite{Sing2015WASP31b} and \cite{Stevenson2016}, where \cite{Stevenson2016} found a weak H\textsubscript{2}O -- J index (0.86 $\pm$ 0.48). Additionally, the retrieved C/O ratio with our Magellan/IMACS data of 0.52$^{+0.87}_{-0.32}$, which is within uncertainties of the retrieved C/O ratios using the other data subsets, suggest an oxygen-rich atmosphere. \added{The C/O ratio is relatively unconstrained, because the only observations that could constrain CO features are two, wide wavelength data points from \textit{Spitzer}. None-the-less} this ratio and the relatively low temperature retrieved (T\textsubscript{p} = 1220$^{+260}_{-250}$ K) imply that water would be a large absorber in WASP-31b's atmosphere \citep{Madhusudhan2012}. Thus, the lack of significant detection of water is likely due to high-altitude clouds muting the water features. 

\subsection{Final Interpretation} \label{sec:Finale}
\indent The possible extremely-muted water features, lack of optical absorption features from K and Na, and retrieved low pressures with all data subsets (see Sections \ref{sec:atmo_retriev_CombMag} \& \ref{sec:atmo_retriev_Previous}) point towards WASP-31b possessing high-altitude clouds. \added{Furthermore, comparing our spectra and retrieval results with atmospheric modeling by \cite{Wakeford2015} and \cite{Pinhas2017} the data suggests: 1) Their likely is a large number of particles of modal size between $\sim$3$\times$10$^{-2}$ and .25$\mu$m lofted high in WASP-31b's atmosphere. 2) The composition of these particles are predominately enstatite and other silicates. 3) These high opacity structures are clouds, which is an accumulation of particles that condense in equilibrium or near equilibrium chemistry; rather than hazes, which form from photochemistry or other non-equilibrium chemical processes. Given that this planet likely has clouds obscuring the optical and near-IR features, spectral observations in the IR (6-30 $\mu$m) could clarify the cloud composition and particle size distribution. As suggested by \cite{Wakeford2015} and \cite{Pinhas2017}, JWST observations would be optimal for such clarity.} 


\section{Summary \& Conclusion} \label{sec:conclusion}

We observed four transit epochs of WASP-31b between 2013 and 2019 with the Magellan/IMACS spectrograph. From these observations we derived optical transmission spectra of the planet between 0.4$\mu$m and 0.97$\mu$m. We excluded the first transit (UT130226), which had inconsistent retrieved parameters,
and combined the other three transits to produce a transmission spectrum with an average precision of 0.0033$R_p/R_s$ per 187{\AA} bin widths. We combined our transmission spectrum with literature HST/WFC3 and \textit{Spitzer} observations and modelled the full datasets using two independent retrieval codes. We find signs of a Rayleigh scattering slope and water features, but neither are significantly detected. We also find no Na or K signals likely due to a high-altitude cloud deck. Our \texttt{PLATON} scattering slope model had the highest $\Delta \ln Z$, and retrieved a Rayleigh scattering slope of 5.3$^{+2.9}_{-3.1}$, log metallicity of 1.3$^{+1.1}_{-1.4}$,  C/O of 0.52$^{+0.87}_{-0.32}$, and cloud top pressure of -3.6$^{+2.7}_{-2.1}$ bars. 

In addition, we ran the retrievals against data from the literature \citep{Sing2015WASP31b,Gibson2017WASP31}, which all showed weak signs of water absorption likely muted by high-altitude clouds. The retrieved scattering slopes for the datasets including the bluer STIS observations were steeper than Rayleigh by over 2$\sigma$. This could mean that WASP-31b's scattering slopes is beyond Rayleigh, but we acknowledge there could be other causes of this retrieved slope. Of all available data on WASP-31b, only the original STIS transit showed strong K absorption. 

From this work we find that the atmosphere of WASP-31b is likely to be significantly covered by high-altitude clouds, explaining high uncertainties of parameters and low significance of feature detection. 

It is imperative to understand which planets form high-altitude clouds or hazes, and how. Understanding this will be key in the JWST era. We need a large library of thoroughly analyzed optical to IR transmission spectra for this. In particular, the optical component allows for absolute abundances of molecules retrieved in the IR by constraining the level at which the molecular features are muted by clouds. Such a library would allow for better comparative studies on observed spectral features and other system parameters (e.g., stellar irradiation levels, spectral type, mass, radius, planet density etc.) to find correlations, and will drastically enhance the field of exoplanetology. Uniform datasets for a wide range of planets, such as the ones ACCESS is building, will be crucial for such analyses.

\acknowledgments 
We appreciate the anonymous referee for helpful comments and feedback. The results reported herein benefited from support, collaborations and information exchange within NASA's Nexus for Exoplanet System Science (NExSS), a research coordination network sponsored by NASA's Science Mission Directorate. This paper includes data gathered with the 6.5 meter Magellan Telescopes located at Las Campanas Observatory, Chile. We thank the staff at the Magellan Telescopes and Las Campanas Observatory for their ongoing input and support to make the ACCESS observations presented in this work possible. We also appreciate the support from the NSF Graduate Research Fellowship (GRFP), grant number DGE1745303. 
B.V.R. thanks the Heising-Simons Foundation for support.
G.W.H. acknowledges long-term support from NASA, NSF, Tennessee State University, and the State of Tennessee through its Centers of Excellence program.
A.J.\ acknowledges support from FONDECYT project 1171208, and from ANID – Millennium Science Initiative – ICN12\_009.
C.D.M. appreciates the help Josh Speagle, a colleague and friend, provided in better understanding the Bayesian statistics behind the detrending and sampling techniques used.

\software{Astropy \citep{astropy2013}, corner \citep{corner2016}, Matplotlib \citep{matplotlib2007}, NumPy \citep{numpy2006}, Multinest \citep{multinest2009}, PyMultiNest \citep{2014BuchnerPyMultiNest}, SciPy \citep{scipy2001}, batman \citep{Kreidberg2015_batman}, george \citep{Mackey2014_george}} dynesty \citep{Speagle_2019Dynesty}, PLATON \citep{Zhang_2019PLATON}, ld-exosim \citep{Espinoza2016_LDlaw}

\facilities{Magellan:Baade, Smithsonian Institution High Performance Cluster
(SI/HPC)}

\bibliographystyle{yahapj}
\bibliography{Paper}

\appendix 

\section{Light Curves} \label{Appx:Lgt_curves}
\begin{figure*}[h!]
    \centering
    \includegraphics[height=1.2\columnwidth]{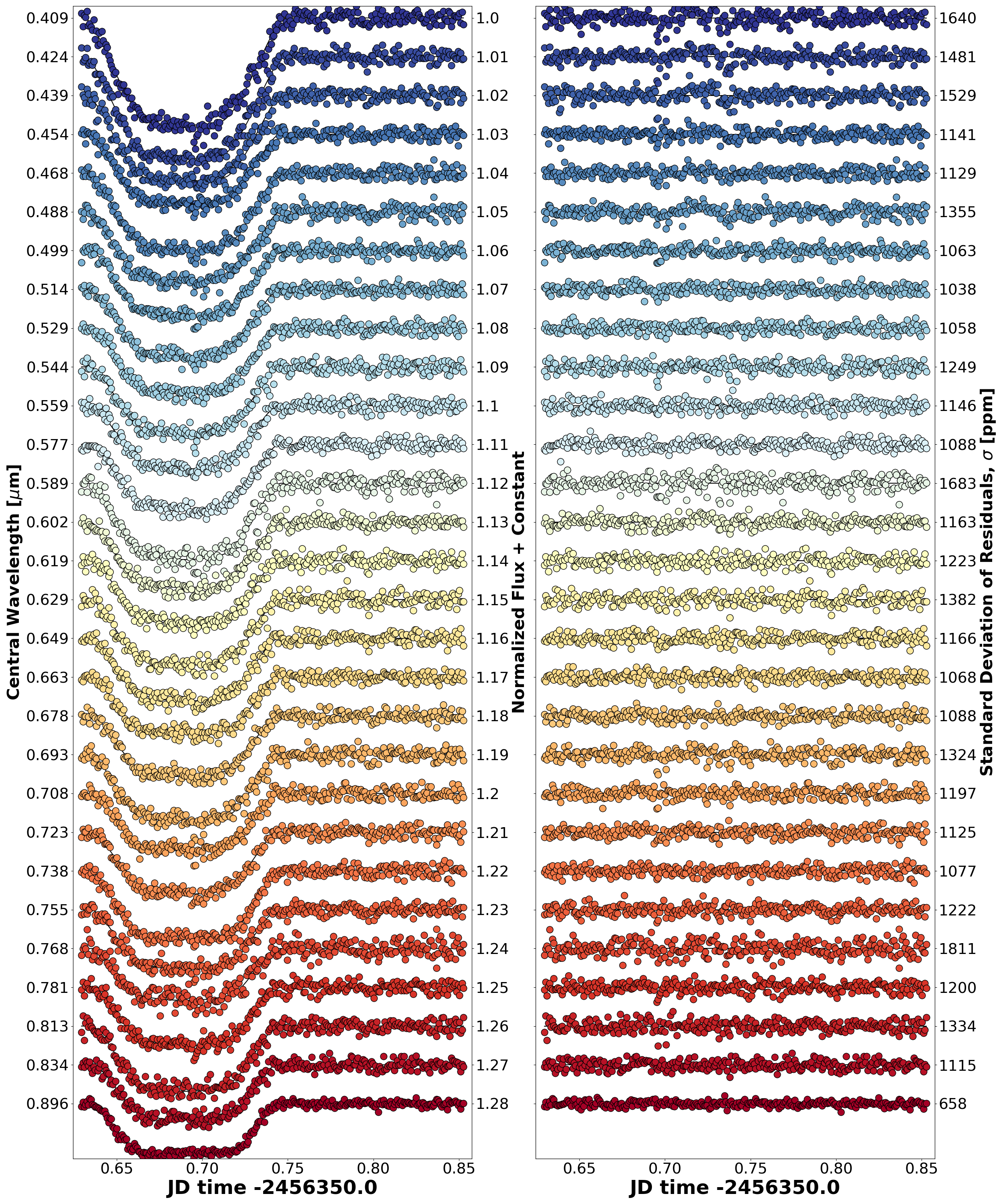}
    \caption{Detrended, binned light curves of transit UT130226 (left) with residuals of the detrended data from the best-fit transit model on the right. The first axis on the left panel displays the mean wavelength of that bin in microns, and the second axis displays the light curves' normalized flux, which were incrementally offset by 0.01 for visual purposes. The axis on the right panel displays the standard deviation ($\sigma$) of the residuals in ppm. Each bin was detrended with model averaging of PCA and GP machine learning techniques. These light curves encompass the spectrum from 4018{\AA} to 9450{\AA}, aside from the spaces caused by chip gaps.}
    \label{fig:binnedLC1} 
\end{figure*}
\begin{figure*}[h!]
    \centering
    \includegraphics[height=1.2\columnwidth]{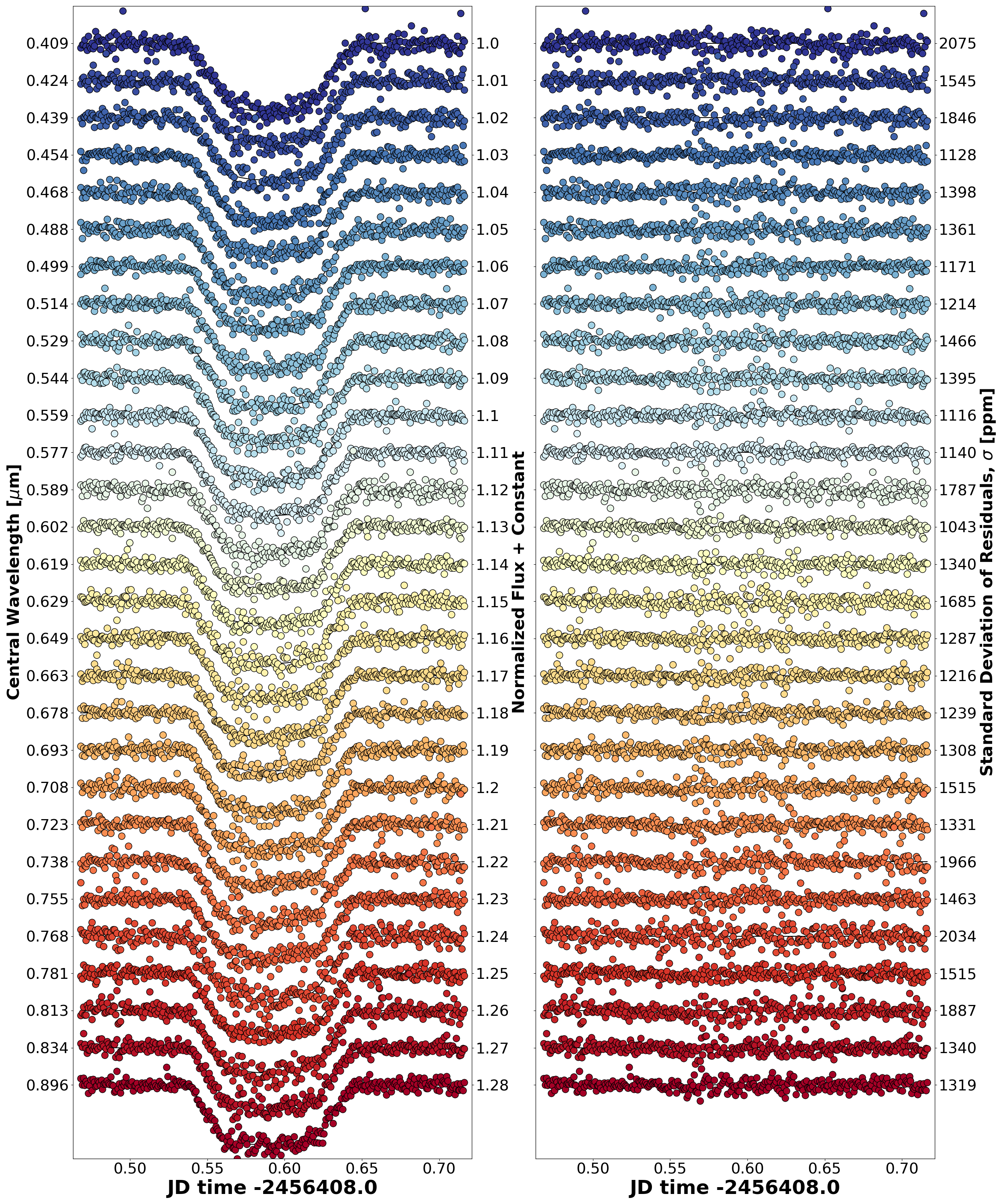}
    \caption{Same as Figure \ref{fig:binnedLC1} but for transit UT130425.}
    \label{fig:binnedLC2} 
\end{figure*}
\begin{figure*}[h!]
    \centering
    \includegraphics[height=1.2\columnwidth]{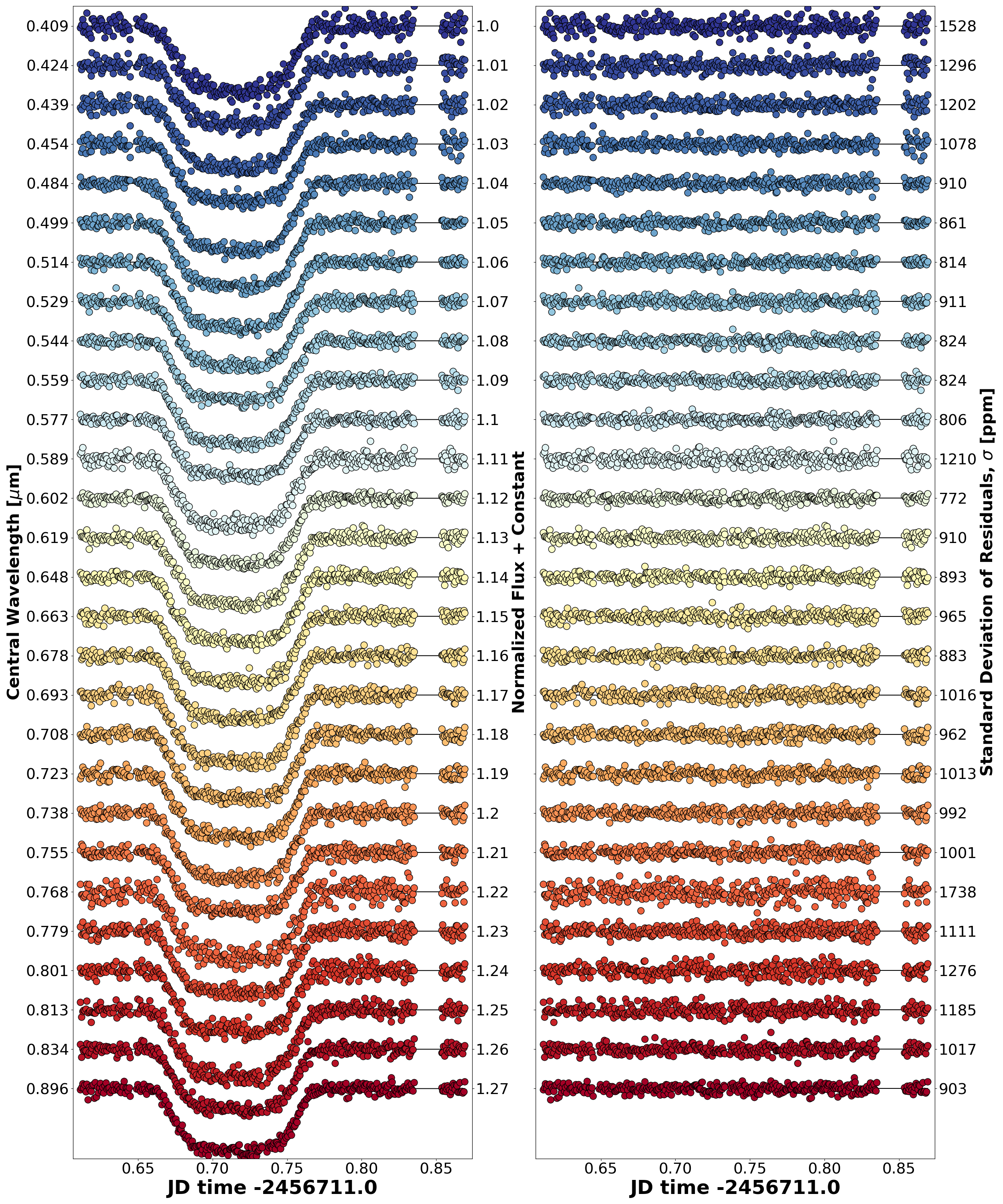}
    \caption{Same as Figure \ref{fig:binnedLC1} but for transit UT140222.}
    \label{fig:binnedLC3} 
\end{figure*}
\begin{figure*}[h!]
    \centering
    \includegraphics[height=1.2\columnwidth]{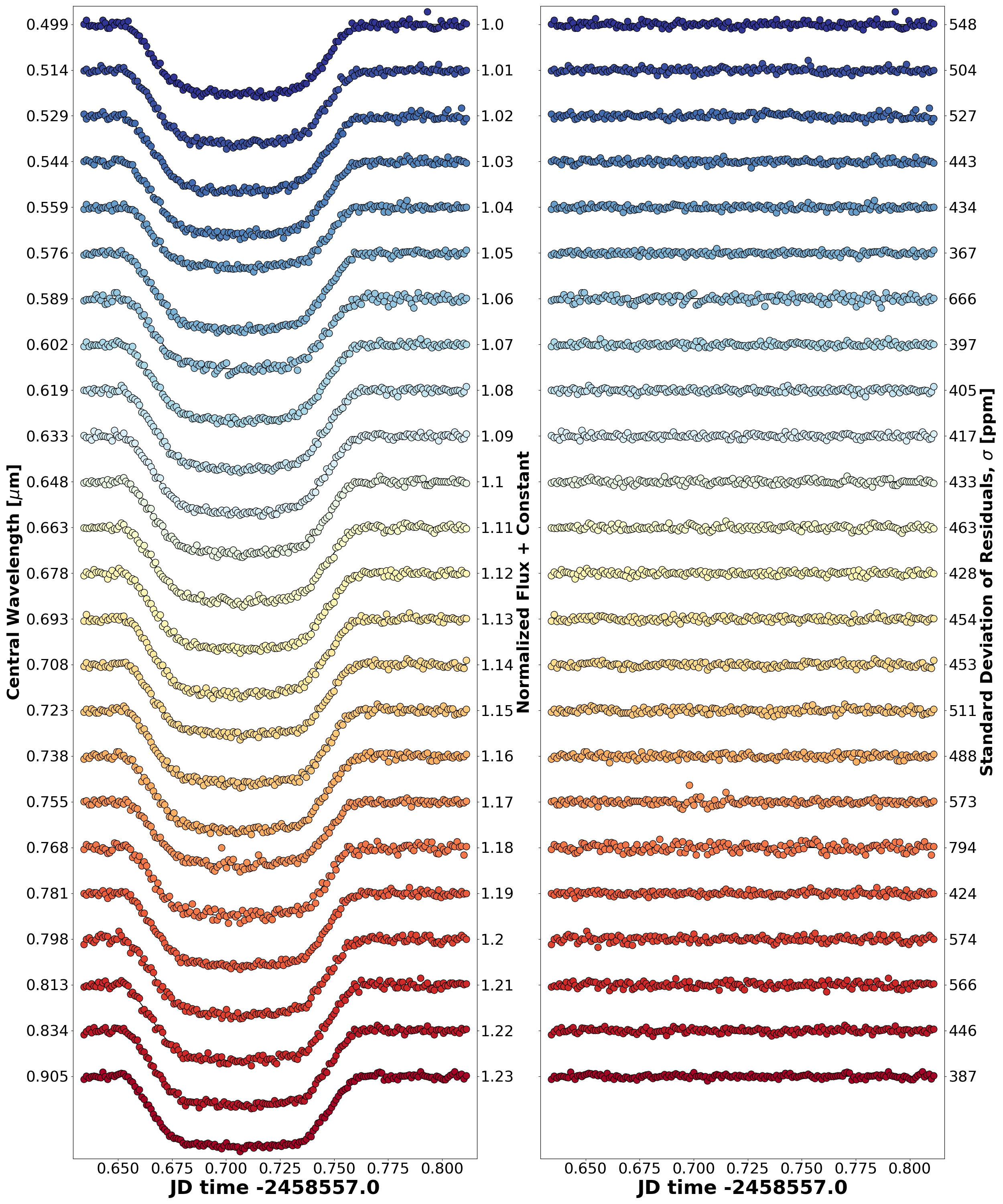}
    \caption{Same as Figure \ref{fig:binnedLC1} but for transit UT190314. Because of the IMACS chip gaps, this particular dataset fully encompasses the spectrum from 4918{\AA} to 9670{\AA}.}
    \label{fig:binnedLC4} 
\end{figure*}

\begin{deluxetable}{|C|C|C|C|C|}[htb]
    \caption{Magellan/IMACS optical transmission spectrum (planet radius/star radius, \textit{p}) for each of our four transit epochs: UT130226, UT130425, UT140222, and UT190314, respectively. The data were produced implementing the reduction and detrending processes discussed in Section \ref{sec:data_red_lc_analysis}. These depths do not include the offsets for combining each spectra. Gaps in the spectra (see Figure \ref{fig:finExtSpec}, prevented a few bins from exactly overlapping in wavelength space. Those bins were still weighted averaged together (x and y direction), with the resulting bin composing of the full wavelength width of all bins. In this table, each bin for the combined transmission spectrum is separated by a vertical line.}
    \label{tab:Trans_Spec_Each}
    \tablehead{\textbf{Wavelength} & \textbf{UT130226} & \textbf{UT130425} & \textbf{UT140222} & \textbf{UT190314} \\
    \textbf{range}~(\si{\angstrom}) & \textbf{R\textsubscript{p}/R\textsubscript{s}} & \textbf{R\textsubscript{p}/R\textsubscript{s}} & \textbf{R\textsubscript{p}/R\textsubscript{s}} & \textbf{R\textsubscript{p}/R\textsubscript{s}}} 
    \startdata
    4018.0-4168.0 & 0.1673^{+0.0071}_{-0.0076} & 0.1361\pm0.0045 & 0.1292^{+0.0047}_{-0.0065} &  ----\\ \hline
    4168.0-4318.0 & 0.1632^{+0.0063}_{-0.0066} & 0.1315^{+0.0043}_{-0.0039} & 0.1216^{+0.0061}_{-0.0071} & ---- \\ \hline
    4318.0-4468.0 & 0.1511^{+0.0067}_{-0.0062} & 0.1305^{+0.0031}_{-0.0029} & 0.1268^{+0.0049}_{-0.0052} & ---- \\ \hline
    4468.0-4618.0 & 0.1347^{+0.0042}_{-0.0043} & 0.1341^{+0.0024}_{-0.0023} & 0.1203^{+0.0051}_{-0.0055} & ---- \\ \hline
    4618.0-4733.0 & 0.1454^{+0.0064}_{-0.006} & 0.1299^{+0.0022}_{-0.002} & ---- &  ---- \\ \hline
    4768.0-4918.0 & ---- & ---- & 0.1302^{+0.0033}_{-0.0037} & ----\\ 
    4835.0-4918.0 & 0.1351^{+0.005}_{-0.0062} & 0.1343^{+0.0029}_{-0.003} & ---- & ---- \\ \hline
    4918.0-5068.0 & 0.1312^{+0.0032}_{-0.0035} & 0.1287\pm0.0019 & 0.1267^{+0.0031}_{-0.0046} &  0.1302^{+0.0044}_{-0.0049} \\ \hline
    5068.0-5218.0 & 0.1329^{+0.0046}_{-0.0047} & 0.1323^{+0.0028}_{-0.0029} & 0.1293^{+0.0028}_{-0.003} &  0.1312\pm0.0023 \\ \hline
    5218.0-5368.0 & 0.1329^{+0.0036}_{-0.0039} & 0.1338^{+0.0033}_{-0.0036} & 0.1301^{+0.0043}_{-0.0056} & 0.1308^{+0.0027}_{-0.002}  \\ \hline
    5368.0-5518.0 & 0.1318^{+0.0043}_{-0.0047} & 0.1314^{+0.0026}_{-0.0027} & 0.1233^{+0.0037}_{-0.0039} & 0.1293^{+0.0032}_{-0.0033} \\ \hline
    5518.0-5668.0 & 0.1284\pm0.0041 & 0.1337^{+0.0016}_{-0.0017} &  0.131^{+0.0014}_{-0.002} & 0.1222\pm0.0051 \\ \hline
    5668.0-5862.9 & 0.1303^{+0.0025}_{-0.0024} & 0.1315^{+0.0018}_{-0.0019} & 0.1202^{+0.0047}_{-0.0045} & 0.1323\pm0.0018 \\ \hline
    5862.9-5922.9 & 0.141^{+0.0065}_{-0.0082} & 0.132\pm0.0021 &  0.1306^{+0.0024}_{-0.003} & 0.1254^{+0.0032}_{-0.0034} \\ \hline
    5922.9-6118.0 & 0.1295^{+0.0073}_{-0.0062} & 0.1295^{+0.0019}_{-0.0017} & 0.1286^{+0.0014}_{-0.0016} &  0.1297^{+0.0019}_{-0.0018} \\ \hline
    6118.0-6256.0 & 0.1286^{+0.0033}_{-0.0037} & 0.1295^{+0.0021}_{-0.0019} & 0.1309^{+0.002}_{-0.0019} &  0.1317^{+0.003}_{-0.0031} \\ \hline
    6256.0-6333.0 & 0.133^{+0.0046}_{-0.0054} & 0.1307^{+0.0031}_{-0.003} & ---- &  ---- \\ 
    6256.0-6406.0 &---- & ---- &  ---- & 0.1316^{+0.0028}_{-0.0029} \\ \hline
    6406.0-6556.0 & ---- & ---- & 0.1293^{+0.0014}_{-0.0017} & 0.1265^{+0.0024}_{-0.0026} \\ 
    6426.0-6556.0 & 0.1282^{+0.0045}_{-0.0042} & 0.1285^{+0.002}_{-0.0019} &  ---- & ---- \\ \hline
    6556.0-6706.0 & 0.1218^{+0.0039}_{-0.0049} & 0.1295^{+0.0023}_{-0.0022} & 0.1302\pm0.0013 & 0.1322^{+0.0028}_{-0.003}\\ \hline
    6706.0-6856.0 & 0.1261^{+0.0035}_{-0.0041} & 0.1264\pm0.0015 & 0.1279^{+0.0018}_{-0.0023} & 0.1299^{+0.0022}_{-0.0023} \\ \hline
    6856.0-7006.0 & 0.1316^{+0.0043}_{-0.005} & 0.1311^{+0.0034}_{-0.0033} & 0.1315^{+0.0017}_{-0.002} & 0.128^{+0.0024}_{-0.0022} \\ \hline
    7006.0-7156.0 & 0.1224^{+0.0039}_{-0.0048} & 0.1301^{+0.0031}_{-0.0029} & 0.1301^{+0.0018}_{-0.0015} & 0.1283^{+0.0026}_{-0.0023} \\ \hline
    7156.0-7306.0 & 0.127^{+0.0046}_{-0.005} &   0.1272^{+0.0016}_{-0.0015} & 0.1278\pm0.0022 & 0.1287^{+0.0016}_{-0.0015}\\ \hline
    7306.0-7456.0 & 0.1322^{+0.0032}_{-0.003} & 0.1276^{+0.0035}_{-0.0033} & 0.1288^{+0.0021}_{-0.002} & 0.1307^{+0.0034}_{-0.0037} \\ \hline
    7456.0-7651.25 & 0.1244^{+0.0042}_{-0.0045} & 0.1278\pm0.0022 & 0.1234^{+0.0035}_{-0.0042} & 0.1221^{+0.0079}_{-0.0121} \\ \hline
    7651.25-7711.25 & 0.1164^{+0.0134}_{-0.008} & 0.1319\pm0.0025 & 0.1312^{+0.003}_{-0.0027} & 0.1195^{+0.006}_{-0.0064} \\ \hline
    7711.25-7860.0 & ---- & ---- & 0.1277^{+0.002}_{-0.0029} & ---- \\
    7711.25-7906.0 & 0.1229^{+0.0037}_{-0.0041} & 0.1276\pm0.0021 & ---- &  0.1259^{+0.0028}_{-0.0026} \\ \hline
    7906.0-8056.0 &  ---- & ---- & ---- & 0.1296^{+0.0028}_{-0.0027} \\ 
    7959.0-8056.0 & ---- & ---- & 0.1256^{+0.0021}_{-0.0024} & ---- \\ \hline
    8056.0-8206.0 & 0.1297^{+0.0041}_{-0.004} & 0.1311^{+0.0023}_{-0.0022} &  0.1315^{+0.002}_{-0.0022} & 0.1331^{+0.0031}_{-0.0039}\\ \hline
    8206.0-8466.0 & 0.119^{+0.0041}_{-0.0047} & 0.128^{+0.0016}_{-0.0015} & 0.1247^{+0.0026}_{-0.0052} & 0.1308\pm0.0016 \\ \hline
    8466.0-9450.0 & 0.1142^{+0.0024}_{-0.0026} & 0.1285^{+0.0024}_{-0.0021} & 0.1269^{+0.0033}_{-0.0052} & ---- \\  
    8466.0-9640.0 & ---- & ---- & ---- & 0.125^{+0.0025}_{-0.0023} \\
    \enddata 
\end{deluxetable}

\begin{deluxetable}{|C|C|C|}[htb]
    \caption{Combined transmission spectrum from transit epochs UT130425, UT140222, and UT190314. The spectra were combined following the procedure outlined in Section \ref{sec:combined_trans_spec}. In the retrieval analysis, because the retrievals do not take asymmetric wavelength errors, the wavelength range of the few overlapping bins were re-centered based on the weighted mean wavelength of the bin.}
    \label{tab:Trans_Spec_Combin}
    \tablehead{\textbf{Wavelength range~(\si{\angstrom})} & \textbf{Mean~(\si{\angstrom})} & \textbf{R\textsubscript{p}/R\textsubscript{s}}}
    \startdata 
    4018.0-4168.0 &4093.00 & 0.1334^{+0.1297}_{-0.1367}\\ \hline 
    4168.0-4318.0 &4243.00 & 0.1288^{+0.1254}_{-0.1323}\\ \hline
    4318.0-4468.0 &4393.00 & 0.1295^{+0.1269}_{-0.1321}\\ \hline
    4468.0-4618.0 &4543.00 & 0.1319^{+0.1297}_{-0.134}\\ \hline
    4618.0-4733.0 &4675.50 & 0.1299^{+0.1279}_{-0.1321}\\ \hline
    4768.0-4918.0 &4868.65 & 0.1326^{+0.1303}_{-0.1348}\\ \hline
    4918.0-5068.0 &4993.00 & 0.1285^{+0.1269}_{-0.1301}\\ \hline
    5068.0-5218.0 &5143.00 & 0.131^{+0.1294}_{-0.1325}\\ \hline
    5218.0-5368.0 &5293.00 & 0.1316^{+0.1299}_{-0.1334}\\ \hline
    5368.0-5518.0 &5443.00 & 0.1289^{+0.1271}_{-0.1307}\\ \hline
    5518.0-5668.0 &5593.00 & 0.1319^{+0.1306}_{-0.1329}\\ \hline
    5668.0-5862.9 &5765.30 & 0.1311^{+0.1298}_{-0.1323}\\ \hline
    5862.9-5922.9 &5892.90 & 0.1303^{+0.1288}_{-0.1317}\\ \hline
    5922.9-6118.0 &6020.45 & 0.1292^{+0.1282}_{-0.1302}\\ \hline
    6118.0-6256.0 &6188.78 & 0.1305^{+0.1292}_{-0.1318}\\ \hline
    6256.0-6406.0 &6294.62 & 0.1312^{+0.1291}_{-0.1333}\\ \hline
    6406.0-6556.0 &6485.00 & 0.1285^{+0.1274}_{-0.1296}\\ \hline
    6556.0-6706.0 &6631.00 & 0.1303^{+0.1293}_{-0.1314}\\ \hline
    6706.0-6856.0 &6781.00 & 0.1276^{+0.1265}_{-0.1286}\\ \hline
    6856.0-7006.0 &6931.00 & 0.1303^{+0.1289}_{-0.1316}\\ \hline
    7006.0-7156.0 &7081.00 & 0.1296^{+0.1285}_{-0.131}\\ \hline
    7156.0-7306.0 &7231.00 & 0.128^{+0.127}_{-0.129}\\ \hline
    7306.0-7456.0 &7381.00 & 0.1289^{+0.1274}_{-0.1305}\\ \hline
    7456.0-7651.25 &7553.62 & 0.1266^{+0.1247}_{-0.1284}\\ \hline
    7651.25-7711.25 &7681.25 & 0.1306^{+0.1288}_{-0.1324}\\ \hline
    7711.25-7906.0 &7798.01 & 0.1272^{+0.1258}_{-0.1285}\\ \hline
    7906.0-8056.0 &7999.69 & 0.1272^{+0.1254}_{-0.1289}\\ \hline
    8056.0-8206.0 &8131.00 & 0.1316^{+0.1301}_{-0.1329}\\ \hline
    8206.0-8466.0 &8336.00 & 0.129^{+0.1279}_{-0.13}\\ \hline
    8466.0-9640.0 &8982.70 & 0.1269^{+0.1254}_{-0.1284}\\ \hline
    \enddata 
\end{deluxetable}

\newpage
\clearpage


\section{Transit UT130226} \label{appx:atmo_retriev_130226}
As can be seen in Figure~\ref{fig:MagTransSpec}, the transit depths of epoch UT130226's transmission spectrum varied widely over our wavelength coverage. Additionally, as can be seen in Figure~\ref{fig:WLC}, there was little baseline before ingress during this observation, which could cause improper detrending of systematics. We tested the impact of our insufficient baseline by comparing the transit duration of epoch UT130226 with those of other epochs when all transit parameters, aside from $p$, were fixed. We also compared the transit depth obtained for this epoch between this test case (with fixed parameters) and our nominal case (with other transit parameters free). For both cases, the transit durations and depths agreed within 1$\sigma$, suggesting that the baseline was sufficient. This is not conclusive support of sufficient baseline, but reasoning to explore if the transmission spectrum is indeed caused by the system. 

We used our retrieval framework to analyze just IMACS transit UT130226 and the IR data to further explore the cause of the outlier transmission spectrum\footnote{The signal of one Magellan/IMACS transit (0.007 $R_p/R_s$) is not enough to detect features like Na or K \citep[$\sim$ 0.006$R_p/R_s$, ][]{Sing2015WASP31b}, but sufficient to interpret the large slope in the spectrum.}. It is possible the transit can be explained by unique physical scenarios in the system during the single epoch, such as the host star being relatively active or the planet having an exceptionally large amount of scattering agents lofted high in the atmosphere. We tested the plausibility of these scenarios by running a variety of retrievals. Table \ref{tab:lnZut130226} shows $\Delta \ln Z$ for these retrievals compared to a featureless spectrum. The evidences for both \texttt{PLATON} and \texttt{Exoretrievals} strongly support stellar activity and a large scattering slope ($\alpha$ = 9.61$^{+0.28}_{-0.46}$ and $\gamma$ = -9.50$^{+0.67}_{-0.35}$). 

The retrieved large scattering slopes are inconsistent with the other Magellan/IMACS and VLT/FORS2 observations (discussed in Sections \ref{sec:atmo_retriev}). Additionally, the best-fit terminator temperatures of 1805$^{+66}_{-126}$ K for \texttt{Exoretrievals} and 1846$^{+35}_{-69}$ K for \texttt{PLATON} were high relative to the other IMACS and FORS2 data ($\sim$1200K) and the calculated equilibrium temperature \citep[1575 K, ][]{Anderson2011_W31discov}. 

An exceptionally high level of stellar activity during the transit of WASP-31b also seems like an improbable scenario, given that our photometeric time series analysis and previous Ca II H \& K spectrospoic analysis \citep{Sing:2016} show signs of low level activity. With \texttt{Exoretrievals} the best-fit stellar activity model was that where $T$\textsubscript{het} = 5720$\pm$310\,K and $f_\mathrm{het}$ = 0.4$^{+0.28}_{-0.20}$, and for \texttt{PLATON} it was where $T$\textsubscript{spot} = 5950$^{+240}_{-380}$ K and $f_\mathrm{spot}$ = 0.31$^{+0.28}_{-0.24}$. These high spot covering fractions imply a completely contrary level of activity from other activity indicators. Thus, we have no concrete explanation of this transmission spectrum's abnormal features, and decided to exclude this transit from our combined data in order to prevent biasing our atmospheric interpretations. 

\begin{deluxetable*}{|l|C|C|C|C|C|C|C|C|}[h!]
    \caption{Same as Table \ref{tab:lnZAllMag} but with the subset of data that included only the UT130226 Magellan/IMACS and IR data. The retrievals with a strong scattering slope ($\gamma\sim-9.5$, $\alpha\sim9.6$) and stellar activity (T\textsubscript{het}$\sim$5700K and $f_{het}\sim$0.4, T\textsubscript{spot}$\sim$5900K and $f_{spot}\sim$0.3) were heavily supported by the \texttt{Exoretrievals} and \texttt{PLATON} evidences, respectively.}
    \label{tab:lnZut130226}
    \tablehead{\multicolumn{6}{|c|}{\bfseries {\large Exoretrievals}} &&\multicolumn{2}{|c|}{\bfseries {\large PLATON}}}
    \startdata 
      \textbf{Model:}  &  \text{featureless}  &  $H_2O$  &  $Na$  &  $K$  &  $H_2O + K +Na$  && \textbf{Model:} &\\ \hline
      featureless &                0.00 &	-0.70 &	-0.60 &	-1.21 &	-0.81 && \text{featureless} & 0.0\\
      scatterers &          	--- &	14.53 &	14.13 &	14.00 &	13.66 & &\text{scattering} & 24.67\\
      activity &           4.08 &	10.04 &	3.02 &	3.40 &	9.45 & &\text{stellar activity} & 12.01\\
      scatterers \& activity & --- &	16.11 &	15.08 &	14.72 &	15.17 & &\text{Both} & 22.93
    \enddata 
\end{deluxetable*}
\newpage
\clearpage

\section{Corner plots} \label{Appx:Corner_plts}
\begin{figure*}[htb]
    \centering
    \includegraphics[height=2\columnwidth]{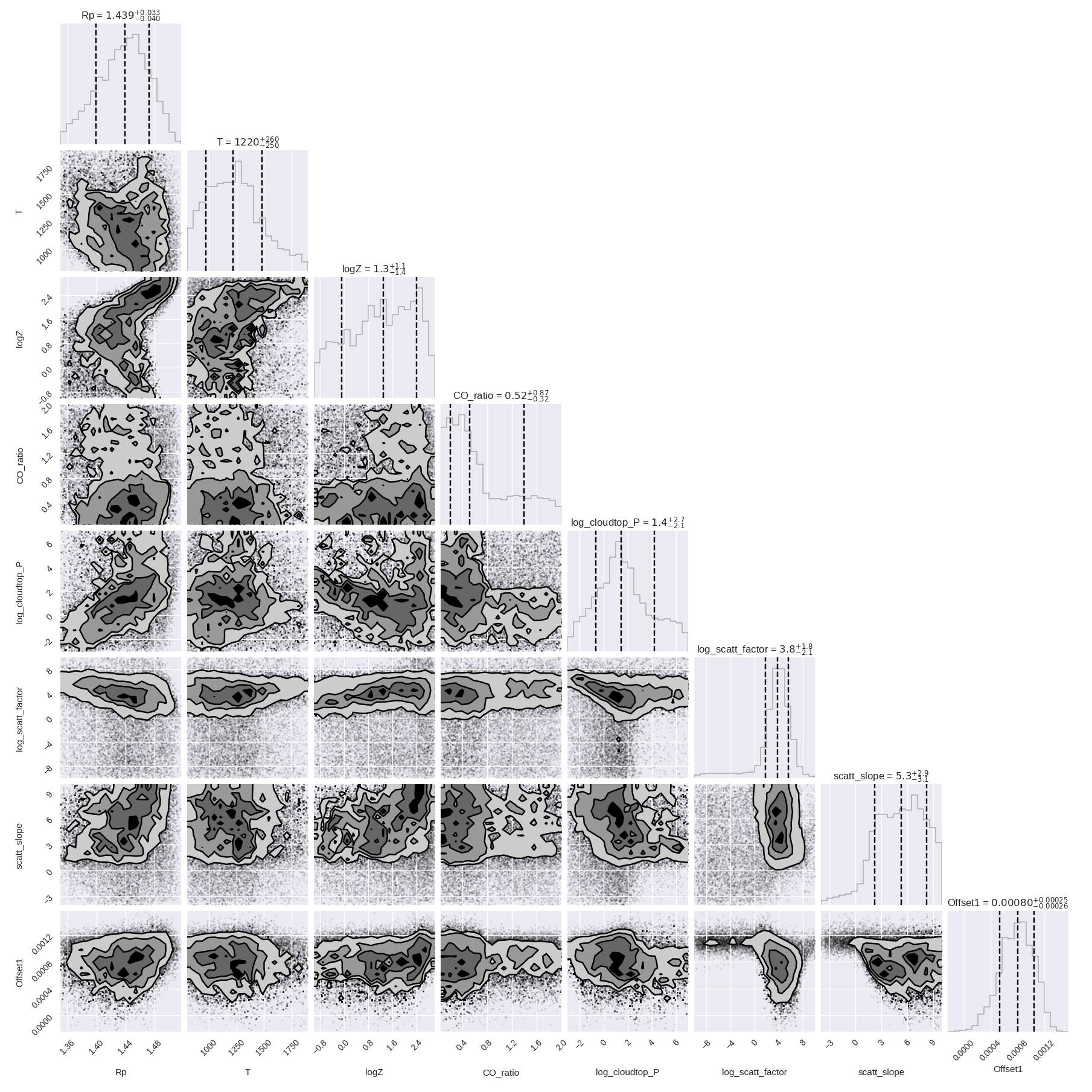}
    \caption{Corner plot of the best-fit retrieval on the combined Magellan/IMACS, HST/WFC3, and \textit{Spitzer} data. This is using the \texttt{PLATON} retrieval and its corresponding transmission spectrum is shown in Figure \ref{fig:PlatIMACS_TranSpec}. Vertical dashed lines mark the 16\% and 84\% quantiles.}
    \label{fig:PlatIMACS_Corn} 
\end{figure*}

\end{document}